\documentclass[aps,prd,superscriptaddress,amsmath,groupedaddress,twocolumn]{revtex4}

\usepackage{graphicx}
\usepackage{dcolumn}
\usepackage{bm}

\newcommand{\BE}{\begin{equation}}
\def\EE{\end{equation}}
\def\BEA{\begin{eqnarray}}
\def\EEA{\end{eqnarray}}

\newcounter{saveeqn}

\begin{document}

\title{Precise $B$, $B_s$ and $B_c$ meson spectroscopy from full lattice QCD}

\author{Eric B. Gregory} 
\affiliation{School of Physics and Astronomy, The Kelvin Building,
       University of Glasgow, Glasgow G12-8QQ, UK}
\affiliation{Department of Physics, University of Cyprus, PO Box 20357, 1678 Nicosia, Cyprus}

\author{Christine T. H. Davies} 
\email[]{c.davies@physics.gla.ac.uk}
\affiliation{School of Physics and Astronomy, The Kelvin Building,
       University of Glasgow, Glasgow G12-8QQ, UK}

\author{Iain D. Kendall} 
\affiliation{School of Physics and Astronomy, The Kelvin Building,
       University of Glasgow, Glasgow G12-8QQ, UK}

\author{Jonna Koponen} 
\affiliation{School of Physics and Astronomy, The Kelvin Building,
       University of Glasgow, Glasgow G12-8QQ, UK}

\author{Kit Wong} 
\affiliation{School of Physics and Astronomy, The Kelvin Building,
       University of Glasgow, Glasgow G12-8QQ, UK}

\author{Eduardo Follana} 
\affiliation{Departamento de F\'{i}sica Te\'{o}rica, Universidad de Zaragoza,
Cl. Pedro Cerbuna 12, E-50009 Zaragoza, Spain}

\author{Elvira G\'{a}miz} 
\affiliation{Department of Physics, University of Illinois, Urbana, 
IL 61801, USA}

\author{G. Peter Lepage} 
\affiliation{Laboratory of Elementary Particle Physics, Cornell University, Ithaca, NY 14853, USA}

\author{Eike H. Muller}
\affiliation{School of Physics, University of Edinburgh, King's Buildings, Edinburgh EH9 3JZ, UK} 

\author{Heechang Na} 
\affiliation{Department of Physics, The Ohio State University, Columbus, OH 43210, USA}

\author{Junko Shigemitsu} 
\affiliation{Department of Physics, The Ohio State University, Columbus, OH 43210, USA}

\collaboration{HPQCD collaboration}

\date{\today}
    
\begin{abstract}
We give the first accurate results for $B$ and $B_s$ meson masses from lattice 
QCD including the effect of $u$, $d$ and $s$ sea quarks, and we improve an 
earlier value for the $B_c$ meson mass. 
By using the Highly Improved Staggered Quark action for $u/d$, $s$ and $c$ 
quarks and NRQCD for the $b$ quarks, we are able to achieve an accuracy in the masses
of around 10 MeV. 
Our results are: $m_B$ = 5.291(18) GeV, $m_{B_s}$ = 5.363(11) GeV and 
$m_{B_c}$ = 6.280(10) GeV. Note that all QCD parameters here are tuned from 
other calculations, so these are parameter free tests of QCD against 
experiment. We also give scalar, $B_{s0}^*$, and axial vector, $B_{s1}$, 
meson masses. We find these to be slightly below threshold 
for decay to $BK$ and $B^*K$ respectively. 

\end{abstract}
   

\maketitle

\section{INTRODUCTION} \label{se:INTRODUCTION}

$B$ meson physics is one of the critical elements of 
the flavor physics programme.
The mass differences between the `heavy' and `light' 
eigenstates of the neutral $B$ and $B_s$ mesons 
are now known experimentally and can be used to 
precisely constrain the ratio of CKM elements 
$|V_{td}|/|V_{ts}|$ if the appropriate theory results
have been calculated with a matching error. A first lattice QCD calculation of 
these mixing matrix elements, including the effect of $u$, $d$ and $s$ sea 
quarks, was given recently and the critical 
quantity $\xi = f_{B_s}\sqrt{B_{B_s}}/f_B\sqrt{B_B}$ 
was obtained to 3\%~\cite{elvirabb}.
Similarly experimental results for the annihilation of charged 
$B$ mesons to leptons via a $W$ boson can be used to 
constrain $V_{ub}$ and $B$ semileptonic decays to $\pi$ 
or $D^{(*)}$ can be used to constrain $V_{ub}$ and $V_{cb}$ 
if the appropriate theory calculations of decay constants 
or form factors are known.  Again calculations of 
these in full lattice QCD have been done and, for example, 
the decay constant of the $B$ meson is obtained to 7\%~\cite{elvirabb}. 

To improve on these lattice QCD results requires 
pinning down and eliminating sources of systematic error 
and testing as stringently as possible that this has been 
done. Here we provide a calculation of the masses 
of $B$ mesons along with their decay constants using 
$b$ quarks in the NRQCD formalism~\cite{lepage:1992tx} with light quarks 
in the Highly Improved Staggered Quark (HISQ) formalism~\cite{Follana:2006rc}. 
The HISQ formalism has improved discretisation errors 
compared to the asqtad improved staggered 
quark formalism used in our previous calculations. 
Because we use NRQCD which can handle hadrons with 
either single or multiple $b$ quarks
we are able to do an
accurate calculation of the $B$ meson masses by linking them 
to meson masses in bottomonium. 
This provides a 
strong test of systematic errors. We are able to handle 
all of the 4 lightest quarks - $u/d$, 
$s$ and 
$c$ -  using the HISQ formalism and are therefore also 
able to calculate mass differences and decay constant
ratios accurately between $B$, $B_s$ and $B_c$ mesons. 

Section~\ref{se:methods} outlines how the Lattice QCD calculation 
was done and Section~\ref{se:results} describes the analysis and results. 
Section~\ref{se:discussion} compares the results to experiment and to 
other lattice QCD calculations. Section~\ref{se:conclusions} gives 
our conclusion.

\section{Lattice QCD calculation}   \label{se:methods}
We are concerned here with mesons with 
one valence $b$ quark and a lighter valence
anti-quark, either $c$, $s$, or $l$. We use the notation $l$, ``light'', 
to refer to either the $u$ or $d$ quark. Everywhere in these 
calculations $m_l=m_u=m_d$, but we will correct for the effects 
of this, along with the effects of missing electromagnetic 
interactions, when we compare to experiment.

The bottom quark moves sufficiently 
slowly inside bound states 
that it is well described by a non-relativistic formulation (NRQCD)
on lattices of moderate lattice spacing. 
The lighter partner quark, with balancing momentum in a meson 
with zero total momentum, is moving 
much faster and requires a relativistic formulation. 
For this we use the HISQ formulation, which offers control 
of discretization errors to the level where even the $c$ quark can be
treated relativistically.

\subsection{The gluon configurations}
 We make use of the MILC collaborations's library of $2+1$-flavor  gauge 
configurations~\cite{Bazavov:2009bb}.
These have two degenerate flavors of light sea quarks and one flavor of 
strange sea quark, formulated with the Asqtad action
\cite{Orginos:1998ue,Orginos:1999cr,Lepage:1998vj}. The 
gluon action is Symanzik-improved through $\cal{O}$$(\alpha_s a^2)$ 
except for terms of $\cal{O}$$(n_f \alpha_s a^2)$ where $n_f$ is the 
number of sea quarks. In fact these terms~\cite{horgan} are of similar 
size to the other $\alpha_s a^2$ terms so in practice the gluon action 
has $\alpha_sa^2$ discretisation errors.  
For this work we use five different ensembles at three different lattice 
spacings, with $a \approx 0.15,0.11,$ and $0.09$ fm. We refer to these
as ``very coarse'', ``coarse'' and ``fine'', respectively.
The configurations have large spatial volumes ($> (2.4\,{\rm fm})^3$).
Table \ref{tab:params} lists the specific ensembles used in this 
work.

\begin{table}
\begin{tabular}{lllllcc}
\hline
\hline
Set & $\beta$ & $a$ (fm) & $au_{0P}m_{l}^{asq}$ & $au_{0P}m_{s}^{asq}$ &  
$L/T$ & $N_{conf}\times N_{t}$ \\
\hline
1 & 6.572 & 0.1583(13) & 0.0097 & 0.0484 & 16/48 & $628 \times 2$\\
2 & 6.586 & 0.1595(14) & 0.0194 & 0.0484 & 16/48 & $628 \times 2$\\
\hline
3 & 6.760 & 0.1247(10) & 0.005 & 0.05 & 24/64 & $507 \times 2$ \\
4 & 6.760 & 0.1264(11) & 0.01 & 0.05 &  20/64 & $589 \times 2 $ \\
\hline
5 & 7.090 & 0.0878(7) & 0.0062 & 0.031 & 28/96 & $530 \times 4$ \\
\hline
\hline
\end{tabular}
\caption{\label{tab:params}Ensembles (sets) of MILC configurations used with 
gauge coupling $\beta$, 
size $L^3 \times T$ and sea 
masses ($\times$ tadpole parameter, $u_{0P}$, taken from the 
average plaquette) 
$m_{l}^{asq}$ and $m_{s}^{asq}$. 
The lattice spacing values in fm are determined using 
the $\eta_s$ meson mass and decay constant~\cite{kendall} 
and given in column 3. 
Column 7
gives the number of configurations and time sources per configuration 
that we used for calculating correlators. On set 5 only half the 
number were used for light quarks.}
\end{table}

In this work we use values of the lattice spacing, $a$, on each 
ensemble determined using the mass and decay constant 
of the pseudoscalar $s\overline{s}$ meson, the 
$\eta_s$. Although this particle is not seen in the real 
world because of mixing with $u\overline{u}$ and $d\overline{d}$ 
which can be prevented on the lattice, its properties 
can be determined from those known from experiment of 
the $\pi$ and $K$ mesons, as described in~\cite{kendall}. 
Table~\ref{tab:params} lists the values obtained in~\cite{kendall}
for the ensembles we are using here. The values of $a$ are larger 
on coarse lattices than those from the more traditional way 
of setting the lattice spacing using the parameter $r_1$, 
but the results agree, as they should, 
in the continuum limit~\cite{kendall}.  

\subsection{HISQ valence quarks}
We use the Highly Improved Staggered Quark (HISQ) formulation~\cite{Follana:2006rc,Follana:2007uv} for valence 
charm, strange and light quarks.

The HISQ action further reduces the residual $\cal{O}$$(\alpha_s a^2)$ 
discretization errors
coming from taste-changing effects
found in the Asqtad formulation. 
It does this with an additional fattening step applied to the gluon 
field coupling to the quarks
\cite{Follana:2006rc,Follana:2007uv}. The errors are reduced by about 
a factor of 3, making HISQ therefore a better action 
to use for $l$ and $s$ quarks. 

We have shown that the HISQ action can even be used for 
$c$ quarks~\cite{Follana:2006rc}, but in that 
case an additional step is needed. 
The `ordinary' tree-level $\mathcal{O}(a^2)$ discretization errors coming 
from the finite difference discretization of the covariant derivative are 
eliminated in both HISQ and Asqtad formulations using an additional 
3-link `Naik' term. The Naik term corrects errors that would 
otherwise appear at $\cal{O}$$(pa)^2$ in the quark, and therefore 
meson, dispersion relation of energy vs momentum. 
A nonperturbative value for the Naik term coefficient (written 
as $1 + \epsilon$) can be 
derived by studying the dispersion relation of the $\eta_c$ 
meson and tuning the coefficient until the square of the 
speed of light in this relation is 1. Here we use the 
values of the Naik coefficient determined in this way in 
~\cite{Follana:2007uv} appropriate to the values 
of $m_ca$ on each ensemble 
\footnote{We have found that the nonperturbative values for 
$\epsilon$ are in fact very close to those 
obtained from a tree level expression in 
terms of $ma$ given in~\cite{newfds}, so in subsequent calculations 
to this one we have simply used the tree level expression}. 
Tuning the Naik coefficient in this way removes all 
discretisation errors from the HISQ action at leading order in 
the square of the velocity of the $c$ quark~\cite{Follana:2006rc}. 
At subleading order in $v_c$ there will be $\cal{O}$$(\alpha_s a^2)$ 
errors. 

On each configuration of the ensembles in Table \ref{tab:params} we have 
generated and stored random-wall source charm, strange and light 
valence propagators. These were calculated in the following way. On the source
time-slice $t_0$ we generate a $U(1)$ vector of random numbers 
$\eta(t_0)_{x^\prime}$. Then we invert to get the HISQ propagator 
$g^{\rm HISQ}({\bf x},t_0)$:
\begin{equation}
g^{\rm HISQ}({\bf x},t_0)=M^{-1}_{x,x^\prime}\eta(t_0)_{x^\prime}.
\end{equation}

For the fine ensemble we use a different random-wall at each of four values
of $t_0$ on every configuration. For the coarse and very coarse configurations
we use two random-wall sources per configuration. Although the time sources 
are equally spaced, their position in time varies from configuration
to configuration through the ensemble
in a pseudo-random manner to further reduce 
auto-correlations within the ensemble.

We use the charm, strange and light HISQ quark propagators
from~\cite{Follana:2007uv}, but note that here we are using a 
different definition of the lattice spacing and this affects 
the meson masses in physical units and therefore the tuning 
of the quark masses. We have included some additional 
quark mass values on the very coarse lattices to be able to correct for 
mistuning. 
We list the HISQ valence parameters used in columns 
$3$ to $7$ in Table \ref{tab:valparams}.
\begin{table}
\begin{tabular}{lllllll}
\hline
\hline
Set & $am_{b}$ & $u_{0L}$ & $am_{c}$ & $1+\epsilon$ & $am_{s}$
 & $am_{0l}^{hisq}$  \\
\hline
1 & 3.4 & 0.8218 & 0.85 & 0.66 & 0.066 & 0.0132 \\
 & 3.6 & 0.8218 & 0.88 & 0.64 & 0.08 &  \\
 & 3.4($c_i \neq 1$) & 0.8218 &  &  &  &  \\
2 & 3.4 & 0.8225 & 0.85 & 0.66 & 0.066 & 0.0264 \\
\hline
3 & 2.8 & 0.8362 & 0.65 & 0.79 & 0.0537 & 0.0067 \\
4 & 2.8 & 0.8359 & 0.66 & 0.79 & 0.05465 & 0.01365 \\
\hline
5 & 1.95 & 0.8541 & 0.43 & 0.885 & 0.0366 & 0.00705 \\
\hline
\hline
\end{tabular}
\caption{\label{tab:valparams} Parameters for the valence quarks. 
$am_{b}$ is the $b$ quark mass in NRQCD, and $u_{0L}$ is the 
tadpole-improvement
parameter used there~\cite{Gray:2005ur}. We use stability 
parameter~\cite{Gray:2005ur} $n$ = 4 in NRQCD 
everywhere. 
Columns 4, 6 and 7 give the charm, strange and light bare quark masses 
for the HISQ action. $1+\epsilon$ is the coefficient of the Naik term
in the charm case~\cite{Follana:2006rc}.  On set 1 we give additional 
values of $m_b$ and $m_s$ that were used for tuning purposes as 
described in the text. We also used alternative $c_i$ coefficients
for the $am_b = 3.4$ case, specifically $c_1=c_6=1.36$ and 
$c_5 = 1.21$, again as described in the text. 
}
\end{table}

The tuning of the HISQ valence $c$ and $s$ masses to their 
correct values on each ensemble is important to avoid mistuning 
effects masquerading as, for example, lattice spacing artefacts 
or sea quark mass dependence. We tune the $c$ mass to give the 
correct $\eta_c$ meson mass on each ensemble. Here the correct 
$\eta_c$ mass has to be adjusted slightly from its experimental 
value of 2.980 GeV to allow for the fact that our lattice 
QCD calculation is happening in a world without electromagnetism 
and without $c$ quarks in the sea and in which we do not 
allow the $\eta_c$ meson to annihilate to gluons (because 
we have not included such `disconnected' pieces in our
$\eta_c$ meson correlators). These effects are all small and 
can be estimated from potential models or perturbation theory. 
We take the physical value of $m_{\eta_c}$ appropriate to 
our calculations to be 2.985(3) GeV~\cite{newfds}, incorporating 
a shift of 2.6 MeV for electromagnetic effects and 
2.4 MeV for annihilation effects. The effect of $c$ quarks 
in the sea is negligible; this will be discussed in 
the next section.  

We tune 
the $s$ quark mass from the value of the mass of the $\eta_s$ 
meson.  This is a fictitious pseudoscalar meson made of $s$ 
quarks, which is not allowed in our lattice calculation to 
mix, by annihilation, with $u\overline{u}$ and $d\overline{d}$ 
mesons. We cannot therefore fix its mass from experiment but 
must do within a lattice QCD calculation, in which we extrapolate 
results for $\pi$, $K$ and $\eta_s$ meson masses and decay 
constants simultaneously to the continuum limit and physical 
point for the $l$ and $s$ quarks. This was done in~\cite{kendall} 
and the value $M_{\eta_s}$ = 0.6858(40) GeV was obtained, as 
the value appropriate to this lattice QCD world. 

The light 
quark valence masses are taken to match approximately the 
light quark masses used in the sea. The way that this is done 
is to take the ratio of the light valence HISQ mass to the appropriate 
HISQ strange mass value to be the same as the ratio of the 
sea light quark mass to its appropriate $s$ mass value (the 
$u_{0P}$ factors cancel in that ratio). This can only be done 
approximately because the sea `strange' quark masses are not 
very well tuned in some cases~\cite{Bazavov:2009bb} and the correct 
value for $s$ is not known very accurately. However, sea quark 
mass dependence is only a very small effect for everything 
calculated here, so this is not a big issue.  

\subsection{NRQCD $b$ quarks}
\label{subse:NRQCD}
One can estimate the velocity of the quarks in bottomonium mesons by 
comparing radial excitation energies to masses. This shows
that in the $\Upsilon$ the $b$ is very non-relativistic with
$v_b^2\approx 0.1$ (in units of $c^2$). By 
comparison, in charmonium $v_c^2\approx 0.3$. In a mixed 
system with a lighter quark the $b$ quark is even slower than in bottomonium. 
Consider that in a $B_c$ meson the reduced quark mass is roughly 1.5 times that
of the $b\overline{b}$ system and 4.5 times that of a $c\overline{c}$ system.
For a constant mean kinetic energy~\cite{schladming} across all three systems, we then 
expect $v_b^2\approx 0.04$ and $v_c^2\approx 0.35$ in the $B_c$. 
For a $b$ quark combined with an even lighter quark, $v_b^2$ will 
get even smaller. Assuming a mean momentum of $\Lambda_{QCD}$ for the 
$b$ quark inside a $B$ or $B_s$ would give $v_b^2 \approx 0.01$. 
An NRQCD approach 
is then well justified for the $b$ quark in $B_c$, $B_s$ and $B_l$ mesons 
when combined with a relativistic approach for the lighter quarks. 

As we have used a random-wall source for the HISQ propagators, it is critical 
that we initialize the NRQCD $b$ propagators $b$ with the 
{\em same} random-wall function $\eta(t_0)_{x^\prime}$ as we used for the 
HISQ propagators. 
This is non-trivial because the HISQ propagators have one only Dirac
component, whereas NRQCD propagators have explicit spin. 
The spin information for the HISQ propagators is tied up with the 
positions of source and sink, however the source site information  
is lost for a random wall source once the propagator 
has been calculated. If we had access to a HISQ propagator from all 
source points to all sink points we could reconstruct the full 
$4\times 4$ spin structure by multiplying by $\Omega(x)\Omega^{\dag}(y)$ 
where $\Omega$ is the staggering operator, given as a product 
of Dirac gamma matrices as: 
\begin{equation}
\Omega(x)=\gamma_1^{x_1}\gamma_2^{x_2}\gamma_3^{x_3}\gamma_t^{x_t}.
\end{equation}
Here we cannot apply the source $\Omega$ to the HISQ propagators, 
but instead we can apply it as a source for the NRQCD propagators. 
This then effectively 
`undoes' the staggering transformation and gives a naive quark source 
that can be combined with a $b$ 
quark source, but it is done after the 
staggered propagators have been made. The $b$-quark source then 
has to have 4 spin components, rather than the usual two for NRQCD 
because we cannot separate the 
upper and lower components of the naive quark source after the fact. 
Our $b$ quark source is constructed from $\Omega(x)$ 
multiplying the appropriate random 
noise at each site, $\eta(t_0)_{x^\prime}$, updating 
the standard method of combining staggered quarks with other formalisms 
that have explicit spin components derived in~\cite{wingate}.  

A further issue is that we must project onto pseudoscalar or vector 
heavy-light mesons directly at the source of the $b$ quark propagator 
rather than combining appropriate spin components at the end. 
We then have to calculate separate $b$ quark propagators for each of 
the pseudoscalar and vector mesons. 

Finally, to enhance our ability to isolate the ground-state behavior, we
smear the $b$ propagator source with a smearing function, $S$, of various
functional forms and differing radii, $r_i$. 
Combining all of these factors we therefore, 
on timeslice $t_0$, initialize the NRQCD
propagator as:
\begin{equation}
G^{\rm NRQCD}_i({\bf x},t_0) = \sum_{x^\prime}S(\left|x-x^\prime\right|;r_i)\eta_{x^\prime}(t_0)\Gamma\Omega(x^\prime),
\end{equation}
where $\Gamma$ is an element of the Dirac algebra chosen to project out either 
a pseudoscalar or a vector heavy-light meson state.  

On subsequent timeslices we evolve the NRQCD propagator recursively in
the standard way~\cite{oldups}.
Note that here, because upper and lower components 
do not mix in NRQCD, the upper and lower halves of our b-quark 
source are simply 
evolved separately with the same evolution equation. 
Our b-quark propagators then  
have only 2 spin components at the sink end. 
The evolution equation is:
\begin{eqnarray}
G_i(x, t+1) &=& \left(1-\frac{\delta H}{2}\right)\left(1-\frac{H_0}{2n}\right)^nU_t
^\dagger(x)\nonumber \\ & & \left(1-\frac{H_0}{2n}\right)^n\left(1-\frac{\delta H}{2}\right)G_i(x
,t),
\end{eqnarray}
with 
\begin{equation}
H_0 = -\frac{\Delta^{(2)}}{2m_b}
\end{equation} 
and 
\begin{eqnarray}
\label{deltaH}
\delta H &=& -c_1\frac{(\Delta ^{(2)})^2}{8(m_b)^3} +  c_2\frac{ig}{8(m_b)^3}
(\Delta \cdot \tilde{E} - \tilde{E} \cdot \Delta) \nonumber \\
&&-c_3\frac{ig}{8(m_b)^3}{\bf\sigma}\cdot({ {\tilde{\Delta} \times \tilde{E}} - 
{\tilde{E}
 \times \tilde{\Delta}}})\nonumber\\
&&-c_4\frac{g}{2m_b}{\sigma}\cdot {\tilde{B}} 
+ c_5\frac{a^2\Delta^{(4)}}{24m_b}
-c_6\frac{a(\Delta^{(2)})^2}{16n(m_b)^2}.
\end{eqnarray}
$\tilde{E}$ and $\tilde{B}$ are improved versions of the
naive lattice chromo-electric and chromo-magnetic fields,
${E}$ and ${B}$. All the gauge fields appearing are 
tadpole-improved by dividing by a tadpole factor, $u_{0L}$, 
derived from the mean link in Landau gauge. 

The equations given above represent the standard NRQCD action, used
in many previous lattice QCD calculations (for example~\cite{oldups} and~\cite{Gray:2005ur}), correct through $v_b^4$.
The largest source of remaining systematic error from this action is 
from radiative corrections to the coefficients of the $v_b^4$ terms 
required to match full QCD through $\cal{O}$$(\alpha_sv_b^4)$
(the $v_b^2$ term is tuned nonperturbatively in fixing the $b$ 
quark mass as described below). 
Here we generally use the tree-level 
values of $c_i=1$ for the constants as we have done before. However, 
we have also done some calculations on set 1 
using values of $c_1$, $c_5$ and $c_6$ 
that include radiative corrections, to be able to gauge the 
size of the systematic error from these terms. 
How the radiative corrections are calculated and further tests 
of them will be described elsewhere~\cite{eike,hpqcdinprep}.
The values we use on set 1 are those appropriate to a 
value of $m_ba$ of 3.4 and $\alpha_s$ in the $V$ scheme 
appropriate to the lattice spacing of that ensemble. 
These are $c_1 = c_6 = 1.36$ and $c_5$ = 1.21~\cite{eike}.  

We list the NRQCD valence $b$ masses used in 
column 2 of Table \ref{tab:valparams} along with the $u_{0L}$ 
parameters. We have used two different masses on set 1, again 
to test for systematics from quark mass tuning. 
Since NRQCD quarks propagate in one direction in time only we improve 
statistics by generating propagators both forwards in time (for $T/2$ 
time units) and backwards in time from each light quark source. 

The $b$ quark mass is tuned by determining $\Upsilon$ and $\eta_b$ meson 
masses~\cite{hpqcdinprep}. Because the zero of energy has been shifted in 
NRQCD we cannot determine meson masses directly from their energy 
at zero momentum as is done in a relativistic formulation. Instead we 
must calculate the `kinetic mass' of a meson, $M$, which 
appears in the relationship 
between $E(p)$ and $p^2$: 
\begin{equation}
E(p) = E(0) + \sqrt{p^2 + M^2} - M.
\label{eq:disp}
\end{equation}
We are able to do this very precisely using random wall sources~\cite{kendall}. 
We determine the kinetic mass of both the $\Upsilon$ and the $\eta_b$ 
mesons and use the spin-average of the two, i.e. 
\begin{equation}
M_{b\overline{b}} = \frac{3M_{\Upsilon} + M_{\eta_b}}{4}
\end{equation}
to tune the $b$ quark mass. The reason that we use the spin-average 
is to avoid systematic errors from the terms in the NRQCD action 
that give rise to spin-splittings. These terms are only included 
at leading ($v_b^4$) order in the action above (equation~\ref{deltaH}), 
and we know that there are sizeable discretisation errors in 
the hyperfine splitting between the $\Upsilon$ and $\eta_b$ 
as a result~\cite{Gray:2005ur}. These errors also have a small effect on the 
kinetic masses~\cite{hpqcdinprep} and we take the spin-average to 
remove them. 

The experimental value for $M_{b\overline{b}}$ that we tune to must be 
adjusted, as described above in the $\eta_c$ 
case, to be the value appropriate to our lattice 
QCD world which is missing some elements of the real world.    
We estimate the effect of the electromagnetic attraction 
of the $b$ and $\overline{b}$ to be 1.6 MeV from a potential model. 
The absence of electromagnetism shifts both $M_{\Upsilon}$ and $M_{\eta_b}$ 
upwards. The shift from not allowing the $\eta_b$ to annihilate 
to gluons we take to be the same as for the $\eta_c$ at 2.4 MeV. 
In addition we must consider the absence of $c$ quarks in the sea. 
This has only a very small effect on charmonium states~\cite{newfds} 
but a somewhat larger effect on bottomonium, because these 
states are more sensitive to exchange of higher momentum gluons 
that can generate a $c\overline{c}$ pair. The effect can be 
estimated perturbatively from that of a massive quark 
loop in the gluon propagator that gives rise to the heavy quark 
potential~\cite{newfds}. It gives rise to a correction to the 
potential which is proportional to a delta function at the 
origin: 
\begin{equation}
V(r) = -\frac{C_f \alpha_s}{r} \rightarrow - C_f \alpha_s \left(\frac{1}{r} + 
\frac{\alpha_s}{10m_c^2} \delta^3(r)\right). 
\end{equation}
This correction is very similar to the hyperfine potential in bottomonium, 
except that it is not spin-dependent and the mass that appears 
in the denominator is $m_c$ and not $m_b$. The hyperfine potential 
induces the mass difference between $M_{\Upsilon}$ and $M_{\eta_b}$ 
of 69 MeV~\cite{exptetab}. 
The potential above has a coefficient 14 times smaller, 
where this factor is given by $[80\pi/(3\alpha_s)]\times(m_c/m_b)^2$ 
and we take $m_b/m_c$ = 4.51 from~\cite{mbmcpaper}. We therefore 
expect a shift of around 5 MeV to both $\Upsilon$ and $\eta_b$. 
Again the absence of $c$ quarks in the sea shifts the mass
values upwards. The experimental $\eta_b$ mass is now 9.391(3) GeV~\cite{pdg} 
and the $\Upsilon$, 9.460 GeV, giving a spin-average of 
9.443(1) GeV. Applying the shifts above with 
50\% errors, we find the appropriate value for $M_{b\overline{b}}$
for us to tune to is 9.450(4)(1) GeV, where the first error is 
from applying the shifts and the second is the experimental error. 

\subsection{HISQ-NRQCD two-point functions}
At the sink end of the propagators we must tie a two-component
NRQCD propagator with the one-component HISQ propagator.
We first convert the HISQ propagator back into a multi-spin object
by multiplying by $\Omega(x)$ at each sink site:
\begin{equation}
G^{\rm HISQ}({\bf x},t)_{ab}=g^{\rm HISQ}({\bf x},t_0)\Omega({\bf x},t)_{ab},
\end{equation}
remembering that the $\Omega^{\dag}$ factor that would normally be 
at the source end has now 
been included in the NRQCD propagators. 

Finally we generate the $B$ meson correlator by combining the 
HISQ propagator and 
NRQCD propagators at the sink timeslice $t$, again with smearing functions and 
appropriate Dirac structure:
\begin{eqnarray}
{C_\Gamma(t-t_0)_{ij}}=\sum_{\bf x}{G^{\rm HISQ}}^\dagger({\bf x},t)\Gamma 
S(\left|x-x^\prime\right|;r_j) \\ \nonumber \times G^{\rm NRQCD}_i({\bf x},t).
\end{eqnarray}
Note that this is a sum over 4 spin components at the source and 
2 at the sink, so $\Gamma$ is either a $2\times 2$ unit matrix for 
the pseudoscalar meson or a Pauli spin matrix for a vector meson. 

\subsection{Extracting physical masses}
We do simultaneous constrained fits~\cite{gplbayes} to the form
\begin{eqnarray}
\label{eq:fitform}
C_{\rm meson}(i,j,t_0;t)&=& \sum^{N_{\exp}}_{k=1} b_{i,k}b^*_{j,k}e^{-E_{k}(t-t_0)
}\\
 &+& \sum^{N_{\exp}-1}_{k^\prime=1}   d_{i,k^\prime}d^*_{j,k^\prime}(-1)^{(t-t_0)
}e
^{-E^\prime_{k^\prime}(t-t_0)},\nonumber
\end{eqnarray}
where $i$ and $j$ index the $3\times 3$ matrix of source and sink smearing 
functions. The second
term fits the oscillating ``parity-partner'' states that appear in most staggered
meson correlators. Our correlation functions cover the range of $t-t_0$ values 
from zero to $T/2$, although we do not fit all the way to $t-t_0 = 0$. Instead 
we 
start at $t_{min} = 2-4$ for $B_s$ and $B_l$ fits and $6-8$ for $B_c$ fits 
to reduce the effect of excited states. 

We constrain the parameters of the fit with prior values and widths, which
are fed into the augmented $\chi^2$ function that the fit minimizes. 
These priors represent very general information about mass splittings 
and amplitudes. The prior value for the ground state mass is simply 
taken from an effective mass plot, with the prior width taken to be a 
factor of 1.5 from this value. 
The mass parameters for the higher mass states enter the fit as 
the logarithm of the mass difference with the state immediately below, 
so that the mass difference is positive and higher mass states remain, 
by definition, higher mass. We take a prior 
value on these mass differences, both for normal states and 
oscillating states, to be $\approx$ 600 MeV, converted to lattice 
units for each fit. The prior width on the mass differences is taken 
to be a factor of 2. The prior on the mass of the lowest oscillating state is 
taken to be $\approx$ 400 MeV above the ground state, with a prior width 
of a factor of 1.5. We use 0.1 $\pm 1.0$ for the prior value and width 
for all amplitudes. Here the prior value of 0.1 is simply to provide a 
non-zero starting point for the fit. The width of 1.0 can be uniform 
across correlators of different smearing functions because we 
normalise them all so that $\sum_x S(x)^2 = 1$ across a timeslice. 

In this way, we are able to obtain high-confidence fits which are stable, both in 
the central value for the ground state mass and 
amplitude and their errors, with respect to
varying the number of exponential functions included, $N_{\exp}$. 
We take our results from fits with $N_{exp}=5$, since all our fits 
are stable by this point.

Where possible (sets 2-4) we simultaneously fit multiple light valence 
channels. That is, by fitting simultaneously $C_{B_s}$ and $C_{B_l}$,
we can eliminate correlated errors from estimations of mass differences
such as $M_{B_s}-M_{B_l}$.

\begin{figure*}
\begin{center}
\includegraphics[width=140mm]{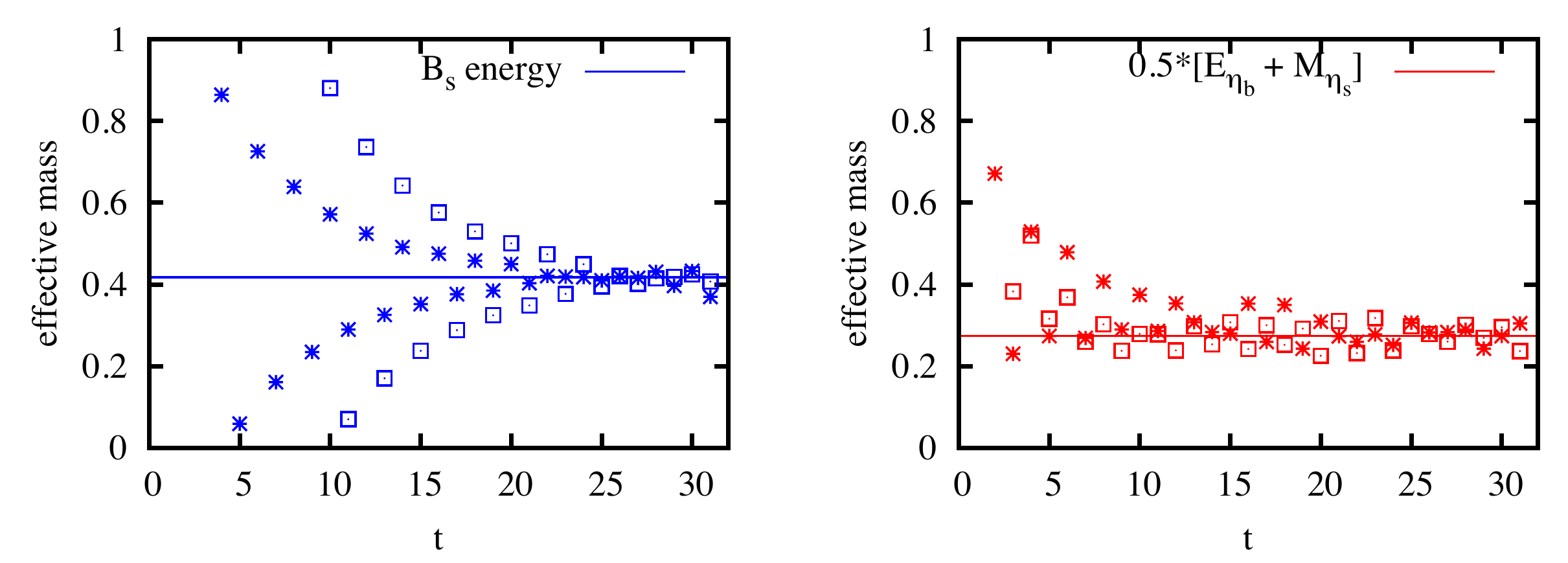}
\end{center}
\caption{The effective mass of the signal (left) and noise (right) 
for $B_s$ correlators on the fine ensemble (set 5). 
The open squares are from correlators with local sources 
and sinks and the bursts from correlators with 
smeared sources and sinks. 
The effective mass is obtained 
from the natural logarithm of the ratio of the 
correlator (or its error) on successive time slices. 
The blue line on the left plot corresponds to the fitted energy of 
the $B_s$ meson and the red line on the right plot to one half the 
sum of that for the $\eta_b$ and $\eta_s$ mesons 
(at the same $am_b$ and $am_s$). 
This figure should be compared to Figure 
3 in~\cite{newfds}. 
}
\label{fig:noise}
\end{figure*}

One important issue with $B$ meson correlators is their 
exponentially falling signal/noise ratio, which means that 
the statistical accuracy that can be obtained on masses 
and decay constants is not as 
high as that of lighter mesons, for example $D/D_s$~\cite{newfds}. 
The variance of the $B_s$ correlator, for example, contains 
$b\overline{b}s\overline{s}$ propagators and can rearrange them 
into an $\eta_b$ and an $\eta_s$. Thus the noise (square root 
of the variance) falls exponentially with a  
lowest energy $(E_{\eta_b}+m_{\eta_s})/2$ at large times
while the signal falls with a lowest energy $E_{B_s}$. This 
means that the signal/noise degrades exponentially with a 
physical energy which is the mass difference between 
the $B_s$ and $(M_{\eta_b} + M_{\eta_s})/2$ (330 MeV). This is illustrated 
in Figure~\ref{fig:noise} where we explicitly compare the effective mass 
of the $B_s$ correlator and the effective mass of its
statistical error, and show that the `mass in the noise' is
as expected. This physical mass difference cannot be altered 
but if we use smearing functions, as we have done here, 
it is possible to extract the ground state $B_s$ mass 
from early $t-t_0$ values, where the noise is less of an issue.  
\footnote{Calculations that use static (infinite mass) $b$ quarks  in 
lattice QCD suffer from an additional problem of zero mass in 
the heavyonium channel that affects the noise, see ~\cite{gpllat91}. 
This has been ameliorated in recent calculations by changing 
the action for the static quark to smear out the quark propagator 
and therefore introduce an effective $E_{\eta_b}$~\cite{sommer}.}

As discussed earlier, the zero of energy is changed in 
the NRQCD formulation 
so the energy
parameters $E_{k}$ and $E^\prime_{k^\prime}$ 
include an energy shift for
which we must correct before comparing to experimental values.
We can do this by comparing the $B$ meson state of 
interest (containing 
1 $b$ quark) to a reference state,
which can also be calculated with NRQCD $b$ quarks on the lattice 
and whose mass is known measured experimentally. That is,
\begin{equation}
M_{B} = (E_{B}-\frac{1}{n}E_{\rm ref})a^{-1} + \frac{1}{n}M_{\rm ref},
\end{equation}
where $E_{\rm ref}$ is calculated on the same lattice ensemble, and $M_{\rm ref}$
comes from experiment (adjusted if necessary for the 
absence of electromagnetism etc. from our calculation). $n$ is the 
number of $b$ quarks in the reference state.
The reference state can also be a linear combination of states,
such as the spin-average of bottomonium states that we will use below. 

To minimize the contribution of the $0.8\%$ uncertainty on $a^{-1}$ 
to the overall uncertainty in $M_{B}$, it is important to choose a 
reference state that makes the quantity $(E_{B}-E_{\rm ref}/n)$ as small as 
possible. We will sometimes do this below by subtracting the masses 
of additional reference states, for example ones made purely of $c$ quarks 
for the $B_c$ case.

\section{Results on the spectrum} \label{se:results}

\subsection{$B_s$ meson mass}
\label{subse:bs}
 To determine $M_{B_s}$ we follow the strategy described earlier, using 
the spin-average of $b\overline{b}$ states as a reference,
and calculate:
\begin{equation}
\label{EXPI}
\Delta_{B_s}= \left(E_{B_s} - \frac{1}{2}E_{b\overline{b}}\right)_{\rm latt}a^{-1}. 
\end{equation}
From this we can reconstruct $M_{B_s}$ using
\begin{equation}
M_{B_s,{\rm latt}} = \Delta_{B_s} + \frac{1}{2}M_{b\overline{b}, {\rm phys}}
\label{eq:mbsrecon}
\end{equation}
Here $E_{b\overline{b}}$ is the spin-average of $\Upsilon$ and 
$\eta_b$ energies at zero momentum calculated with the same 
NRQCD action and on the same configurations as used for calculating the 
$B_s$ meson energies. $M_{b\overline{b}}$ is used to tune the 
$b$ quark mass, as discussed earlier, and $M_{b\overline{b},phys}$ is 
the physical value taken from experiment, but adjusted (to 9.450 GeV) for the 
lattice QCD world (missing electromagnetism, $\eta_b$ annihilation 
and charm-in-the-sea). 
To compare our results for $M_{B_s,latt}$ to experiment 
we have to add corrections to put back in missing electromagnetism and 
charm-in-the-sea effects. These corrections are negligible, 
however, as we will discuss below. 

Table~\ref{tab:bsresults} lists all our fitted values needed for 
determination of the $B_s$ meson mass.  Note that the error 
on the fitted $B_s$ meson energy is larger than any of 
the errors on the other fitted energies. This is because of the 
signal/noise problem in the $B_s$ correlator discussed 
earlier. The other correlators used here do not have that 
problem and the fits give much more precise results for 
ground state masses. Details of these other fits are given 
elsewhere~\cite{kendall}.  

\begin{table*}
\begin{ruledtabular}
\begin{tabular}{llllllllll}
Set & $am_{b}$ & $aM_{b\overline{b}}$ & $aE_{b\overline{b}}$ & $am_s$ & $aM_{\eta_s}$
 & $aE_{B_s}$  & $a\Delta^{hyp}_s$ & $a\Delta^{0^+-0^-}_s$ & $a\Delta^{1^+-1^-}_s$ \\
\hline
1 & 3.4 & 7.260(9) & 0.27843(8) & 0.066 & 0.52524(36) & 0.6409(11) & 0.0343(11) & 0.310(11) & 0.308(14) \\
 & 3.4 &  &  & 0.080 & 0.57828(34) & 0.6539(10) & 0.0349(9) & 0.317(11) & 0.309(12) \\
 & 3.6 & 7.688(5) & 0.27662(7) & 0.066  &  & 0.6466(13) & 0.0324(14) & 0.300(15) & 0.307(16) \\
 & 3.6 &  &  & 0.080  &  &  0.6604(9) & 0.0332(9) & 0.315(11) & 0.308(11) \\
 & 3.4($c_i \neq 1$) & 7.248(4) & 0.28048(7) & & & & & & \\
2 & 3.4 & 7.261(9) & 0.27902(7) & 0.066 & 0.52458(35) & 0.6417(14) & 0.0359(21) & 0.299(17) & 0.316(19) \\
\hline
3 & 2.8 & 5.996(8) & 0.28538(3) & 0.0537 & 0.43118(18) & 0.5470(15) & 0.0287(19) & 0.215(17) & 0.254(6) \\
4 & 2.8 & 5.992(5) & 0.28465(6) & 0.05465 & 0.43675(24) & 0.5527(16) & 0.0261(27) & 0.253(8) & 0.249(16) \\
\hline
5 & 1.95 & 4.288(10) & 0.25985(5) & 0.0366 & 0.30675(12) & 0.4172(10) & 0.0189(12) & 0.1708(48) & 0.166(11) \\
\end{tabular}
\end{ruledtabular}
\caption{\label{tab:bsresults} Results for energies and kinetic masses 
needed for the determination of the mass of the $B_s$ meson.  
All the results are in units of the lattice spacing. 
For each set we list the valence $b$ quark mass and its associated 
kinetic mass and energy for the spin-average of $\Upsilon$ and 
$\eta_b$ states. We also give the valence $s$ quark mass and 
its associated $\eta_s$ meson mass. 
These values are also given in~\cite{kendall}. 
Where we have used multiple $b$ and $s$ masses on set 1, we 
give the $\eta_s$ and $E_{b\overline{b}}$ values only once to 
avoid confusion. 
In column 7, we give the fitted energy of the $B_s$ meson 
(i.e. $E_0$ from fits to the form given in equation~\ref{eq:fitform}). 
In column 8 we give 
the hyperfine splitting, $\Delta^{hyp}_s = E(B_s^*)-E(B_s)$, 
discussed in section~\ref{se:hyp}.
This column is 
largely from~\cite{Gregory:2009hq} but includes some additional 
values on set 1 that we use for studying systematic errors. Note also 
that the value for set 1 on line 1 is different from that in~\cite{Gregory:2009hq} 
although consistent with it. Here we use a value from a fit to 
the $B_s$ and $B_s^*$ correlators alone, rather than from a full fit including
$B_l$ and $B_l^*$, to be in keeping with the other $B_s$ 
values given on set 1.   
Columns 9 and 10 give the values of mass differences between scalar and 
pseudoscalar and between axial vector and vector respectively, 
discussed in section~\ref{se:scalar}. 
}
\end{table*}

\begin{figure}
\begin{center}
\includegraphics[width=80mm]{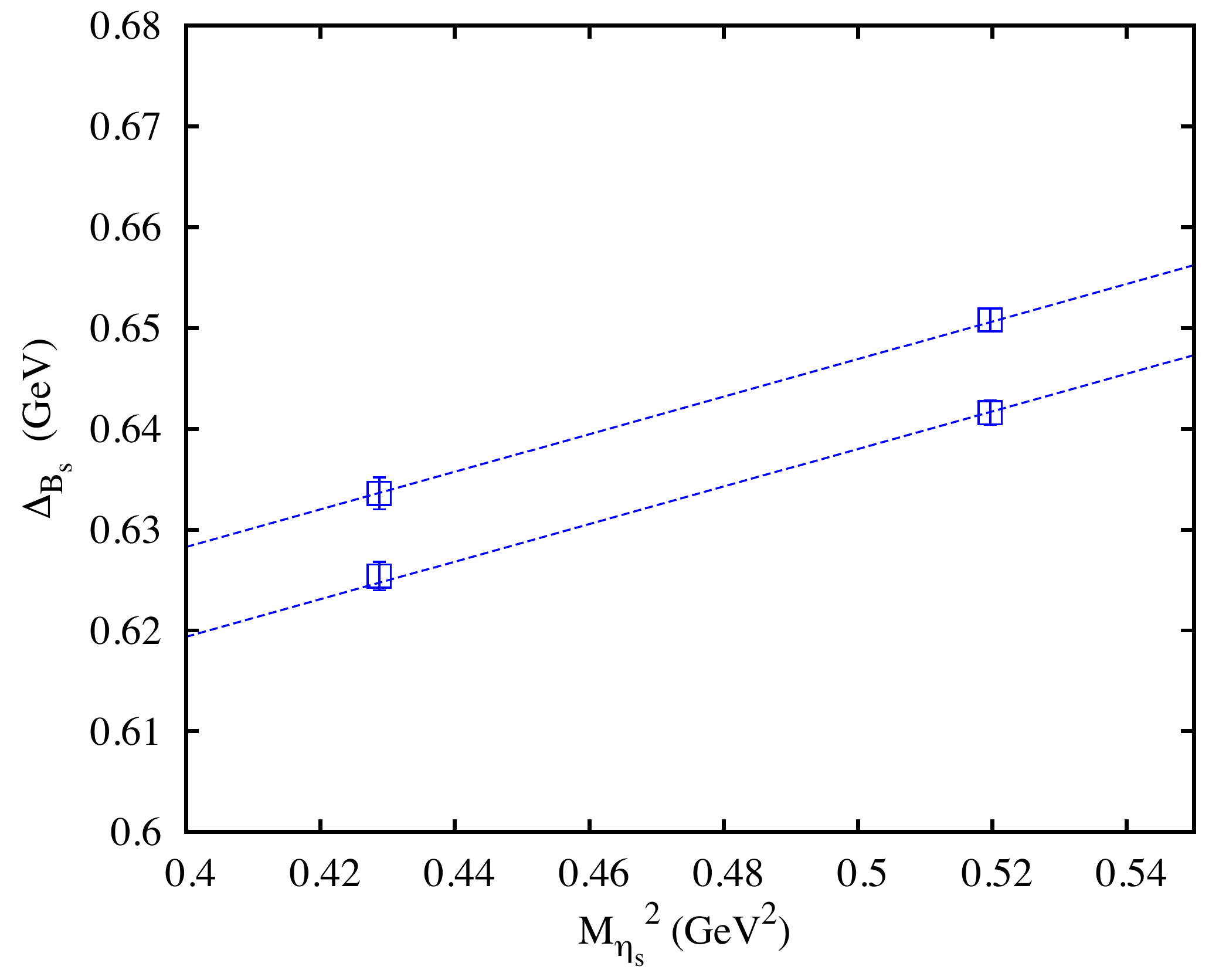}
\end{center}
\caption{Results from set 1 for the mass of the $B_s$ meson 
(specifically the difference between that mass and one 
half of the spin-averaged mass of $\Upsilon$ and $\eta_b$) 
as a function of the square 
of the $\eta_s$ meson mass, acting as a proxy for the 
strange quark mass. 
The errors are statistical only, since lattice spacing errors 
affect all the points together. 
The lines are fits to the results allowing 
linear terms in $M_{\eta_s}^2$ and $M_{b\overline{b}}$. Here the lines 
join points for a fixed $b$ quark mass. See Figure~\ref{fig:mbstuneb}
for the equivalent as a function of $M_{b\overline{b}}$. 
}
\label{fig:mbstunes}
\end{figure}

\begin{figure}
\begin{center}
\includegraphics[width=80mm]{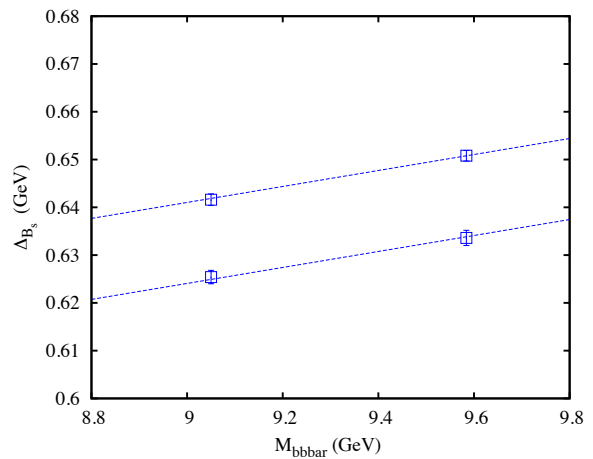}
\end{center}
\caption{Results from set 1 for the mass of the $B_s$ meson 
(specifically the difference between that mass and one 
half of the spin-averaged mass of $\Upsilon$ and $\eta_b$) 
as a function of the spin-averaged kinetic mass of the 
$\Upsilon$ and $\eta_b$, acting as a proxy for the $b$ quark mass. 
The errors are statistical only, since lattice spacing errors 
affect all the points together. 
The lines are fits to the results allowing 
linear terms in $M_{\eta_s}^2$ and $M_{b\overline{b}}$. Here the lines 
join points for a fixed $s$ quark mass. See Figure~\ref{fig:mbstunes}
for the equivalent as a function of $M_{\eta_s}^2$. 
}
\label{fig:mbstuneb}
\end{figure}

In Figures~\ref{fig:mbstunes} and~\ref{fig:mbstuneb} we show how 
$\Delta_{B_s}$ varies with the square of the $\eta_s$ mass 
and the spin-averaged $b\overline{b}$ mass from our results 
on set 1. These results allow us to correct for, and estimate 
the errors from, mistuning $b$ and $s$ masses. 
The lines are simple linear fits in $M_{b\overline{b}}$ and 
$M_{\eta_s}^2$.
The slope of $\Delta_{B_s}$ against $M_{\eta_s}^2$ is 0.19, 
in good agreement with that expected from the experimental 
data comparing $B_s$ and $B$. The slope against $M_{b\overline{b}}$ 
is very small because the $b$ quark 
mass effects naively cancel in $\Delta_{B_s}$. 
However some residual dependence remains and gives a slope of
0.017, somewhat smaller than the experimental result of 0.033 
obtained over a much larger mass range from comparing $B_s$ 
and $D_s$. 

Table~\ref{tab:tune} gives the values of $\Delta_{B_s}$, adjusted 
for mistuning by using the slopes given above and the mismatch 
of $M_{b\overline{b}}$ and $M_{\eta_s}^2$ on each ensemble 
compared to the physical values (9.450 GeV and $(0.6858\, \rm{GeV})^2$ 
respectively). We take a 50\% error on any shift applied 
for mistuning. 
Errors from mistuning are smaller than the statistical 
errors except on sets 1 and 2. Table~\ref{tab:tune} 
also gives the error from the uncertainty in the lattice 
spacing from Table~\ref{tab:params}. The error is smaller 
by a factor of 2 than the naive result of multiplying 
$\Delta_{B_s}$ by the percentage error in $a^{-1}$. The 
reason is that changing the lattice spacing requires 
the masses to be retuned and this affects $\Delta$ in 
the opposite direction.    

\begin{table}[ht]
\begin{tabular}{cccc}
\hline
\hline
Set & $\Delta_{B_s}$ (GeV) & $\delta x_l$ & $\delta x_s$ \\
\hline
1 & 0.6392(16)(28)(25) & 0.15 & -0.08 \\
2 & 0.6382(17)(85)(29) & 0.34 & -0.08  \\
\hline
3 & 0.6401(24)(02)(26) & 0.09 & 0.29  \\
4 & 0.6433(25)(13)(29) & 0.22 & 0.29  \\
\hline
5 & 0.6417(22)(20)(26) & 0.18 & 0.09  \\
\hline
\hline
\end{tabular}
\caption{\label{tab:tune} Results for $\Delta_{B_s}$ 
(the mass difference between the $B_s$ meson 
and the spin average of $\Upsilon$ and $\eta_b$ masses) 
on different ensembles 
after tuning to the correct valence $b$ and $s$ quark masses.
The 3 errors listed are statistical, tuning and from the 
uncertainty in the lattice spacing. Columns 3 and 4 give 
$\delta x_l$ and $\delta x_s$, the fractional mistuning of the 
sea quark masses in units of the $s$ quark mass, as defined 
in the text.}
\end{table}

The resulting errors on the tuned values for $\Delta_{B_s}$ are 
typically less than 1\%, around 4 MeV. Within these uncertainties
we are not able to distinguish any dependence on sea quark 
masses or the lattice spacing. Sea quark mass effects are expected 
to be very small, because the $B_s$ is a gold-plated particle 
and has no valence light quarks. The lattice spacing dependence 
depends on the quantity chosen to fix the lattice spacing. 
Earlier reporting of these results~\cite{ericlat09}, using 
the variable $r_1$ to set the scale, did show visible lattice 
spacing dependence. Here it appears, perhaps not surprisingly, 
 as if $\Delta_{B_s}$ has 
the same discretisation errors as the $\eta_s$ used to the 
fix the scale.  

\begin{figure}
\begin{center}
\includegraphics[width=80mm]{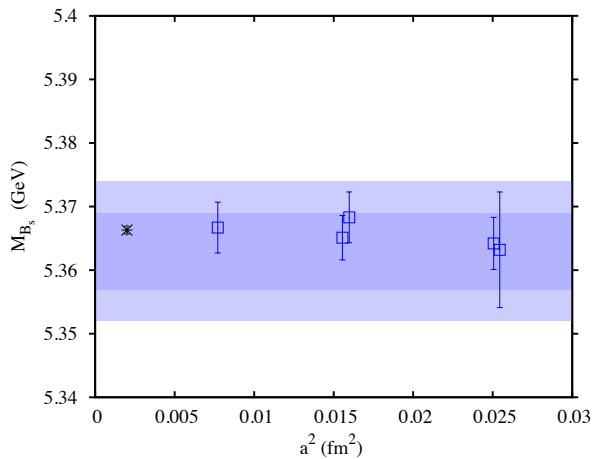}
\end{center}
\caption{Results for the mass of the $B_s$ meson 
tuned to the correct $s$ and $b$ quark masses on 
each ensemble, plotted against the square of 
the lattice spacing. The errors on the points 
include statistical and tuning errors and the 
uncertainty in the lattice spacing. 
The dark shaded band is our physical result, allowing 
for sea quark mass and lattice spacing 
dependence as described in 
the text. 
The width of the light shaded 
band reflects our full error as given in the error 
budget, Table~\ref{tab:errors}. The black star is the 
experimental result~\cite{pdg}, offset from $a^2=0$ for 
clarity.  
}
\label{fig:mbsasq}
\end{figure}

The tuned results from Table~\ref{tab:tune} are used to 
reconstruct $M_{B_s, latt}$ 
(using equation~\ref{eq:mbsrecon}) and this is plotted 
in Figure~\ref{fig:mbsasq} against the square of the lattice 
spacing. In order to quote a physical value that can be 
compared to experiment we need to fit our results as a 
function of lattice spacing and sea quark mass so that 
systematic errors from such dependence can be fed into 
our final error. The sea quark mass dependence we take 
to be a simple polynomial form in the variables $\delta x_s$ 
and $\delta x_l$, defined by 
$\delta x_q = ({m_{q,{\rm sea}}-m_{q,{\rm sea},{\rm phys}}})/{m_{s,{\rm sea},{\rm phys}}}$.   
These variables were used in~\cite{newfds} but must be 
adjusted here consistently for the change in definition 
of the lattice spacing and the new values are given in 
Table~\ref{tab:tune}. Any sea quark mass dependence 
identified in our fit can be extrapolated to the physical 
point where $\delta x_l = \delta x_s =0$, and our errors 
allow for dependence not resolved by our fit. 

The lattice spacing dependence is a trickier issue 
in NRQCD because we cannot extrapolate naively 
to $a=0$. What we need to do is to fit the lattice 
spacing dependence and assess, using information 
from the fit, how much of the dependence is 
physical and how much unphysical, and allow for 
both in the final error. Physical dependence on 
the lattice spacing will arise from discretisation 
errors in the gluon and sea quark actions, and in the light 
valence quark (HISQ) action. We expect this 
dependence to be $\cal{O}$$(\alpha_s a^2)$ at leading 
order as discussed earlier. 

The NRQCD action also has 
discretisation errors. These are corrected at tree level 
by the terms with coefficients $c_5$ and $c_6$ in equation~\ref{deltaH}.  
Beyond tree level $c_5$ and $c_6$ have an expansion in 
powers of $\alpha_s$, required for NRQCD to match QCD at 
that order, whose coefficients depend on $am_b$. This 
dependence will typically be mild for large $am_b$ but 
become singular as $am_b \rightarrow 0$. 
This has been explicitly checked for the $c_5$ and $c_6$ coefficients for 
a slightly different action in~\cite{morning} and 
results have also been derived for this action~\cite{eike, hpqcdinprep}. 
They show almost no $am_b$ dependence for $am_b > 1$. 
In general, however, the coefficients of discretisation 
corrections can be $am_b$ dependent in NRQCD and therefore 
our discretisation errors can be $am_b$ dependent. We need 
to allow for a mild nonsingular dependence (i.e. appropriate 
to the values of $am_b$ that we are using) in our fits, so that the systematic 
error from this can appear in our final results.  
Since any smooth function can be expanded over a limited 
range using a polynomial, we simply allow for linear 
and quadratic terms in the variable $\delta x_m = (m_ba-2.7)/1.45$.
The factors 2.7 and 1.45 are chosen so that 
$\delta x_m$ changes from -0.5 on the fine lattices to 
+0.5 on the very coarse lattices as $m_ba$ covers the range that 
we have used. 

We therefore fit our results for $\Delta_{B_s}$ to the following form: 
\begin{eqnarray}
\label{eq:fitbsextrap}
\Delta_{B_s}(a,\delta x_l, \delta x_s) &=& \Delta_{B_s,{\rm phys}}[1 \\ \nonumber 
&+&\sum_{j=1}^2 c_j(\Lambda a)^{2j}(1 + c_{jb}\delta x_m + c_{jbb}(\delta x_m)^2) \\ \nonumber 
&+& 2b_l\delta x_l(1+c_l(\Lambda a)^2) \\ \nonumber
&+& 2b_s\delta x_s(1+c_s(\Lambda a)^2) \\ \nonumber
&+& 4b_{ll}(\delta x_l)^2 + 2b_{ls}\delta x_l\delta x_s + b_{ss}(\delta x_s)^2].
\end{eqnarray} 
We take the prior on $\Delta_{B_s,{\rm phys}}$ to be 0.6(2). The priors 
on the sea quark mass dependence, $b_l$ and $b_s$, are 
taken to be 0.00(7). Sea quark mass effects are suppressed by a 
factor of 3 over valence mass effects and here valence effects 
correspond to a slope in quark mass of less than 0.2.  The 
priors for the $b$ parameters corresponding to the 
quadratic sea mass dependence 
are then set to $(0.2)^2/3$ i.e. 0.000(13). 
We take the scale of the physical $a$-dependence to be the scale 
of $\Lambda$ = 400 MeV, since we expect it to be set by typical 
internal meson momenta in QCD. 
The coefficients of the quadratic $a$ dependence, $c_1$, $c_l$ and $c_s$, 
should be $\cal{O}$$(\alpha_s)$ so we take priors of 0.0(5). For $c_2$, 
and the $am_b$ dependence of the discretisation errors, $c_{jb}$ and $c_{jbb}$, we take 
a very conservative prior of 0(1). 

The result from the fit is $\Delta_{B_s, {\rm phys}}$ = 0.638(6) GeV. The fit sees 
no dependence on lattice spacing, $am_b$ or sea quark masses but our 
final error allows for all of these. The fit, and its error, is robust 
under changes in the number of fit parameters, for example, including 
or not including the $a^4$ terms in equation~\ref{eq:fitbsextrap}. 
It is also robust under changes the prior widths. For example, doubling 
the prior width on the lattice spacing or sea quark mass dependence 
changes the final result by less than 1 MeV. 

In the error budget in Table~\ref{tab:errors} we separate 
the 6 MeV error into component parts coming from the errors on 
the original 
data points (statistics, tuning and uncertainty in the lattice 
spacing) and the errors coming from the lattice spacing and 
sea quark mass dependence of the results, using the method 
described in~\cite{newfds}. The error on the 
original data points dominates. 

To reconstruct $M_{B_s, {\rm latt}}$ we must add 9.450/2 GeV to $\Delta_{B_s,{\rm phys}}$
as in equation~\ref{eq:mbsrecon}. This gives $M_{B_s}$ in the 
lattice world with no electromagnetism or $c$ quarks in the sea. The latter
effect should be negligible for the $B_s$ since it is a much 
larger particle than the $\Upsilon$ or $\eta_b$ and therefore 
much less sensitive to the gluon exchange that could create 
a $c\overline{c}$ pair. The effect of electromagnetism can be 
estimated following~\cite{newfds}. There we gave a phenomenological 
formula for electromagnetic and $m_u/m_d$ mass difference 
effects in heavy-light mesons: 
\begin{equation}
M(Q,q) = M_{{\rm sim}}(Q,q) + A e_qe_Q + B e_q^2 + C(m_q-m_l).
\label{eq:em}
\end{equation}
$M_{{\rm sim}}$ is the mass of the meson in the absence of 
electromagnetism and with $m_u = m_d$. From experimental 
charged and neutral $B$ and $D$ meson masses we 
determined $A \approx$ 4 MeV and $B \approx$ 3 MeV. 
For the $B_s$ then this formula gives a shift between 
$M(Q,q)$ and $M_{{\rm sim}}(Q,q)$ as a result of electromagnetism 
of -0.1 MeV, a very small 
effect. We make no correction for this, but add an error 
for it to our error budget. 

\begin{table}[ht]
\begin{tabular}{lccc}
\hline
\hline
Error & $M_{B_s}$ & $M_{B_c, hh}$ & $M_{B_c, hs}$ \\
\hline
Stats/tuning/uncty in $a$ & 5.5 & 2.9 & 14.0 \\
Lattice spacing dependence & 0.5 & 2.9 & 8.0 \\
$m_{q,sea}$ dependence & 3.0 & 1.0 &  4.0 \\
spin-ind. NRQCD systs. & 8.0 & 6.0 & 7.5\\
spin-dep. NRQCD systs. & 3.5 & 4.0 & 1.0 \\
uncty in $M_{\eta_s}$ & 1.0 & - & 2.3\\
em, annihiln, $c_{sea}$ in $b\overline{b}$ & 2.5 & $2.5^*$ & 0.0\\
em, annihiln, $c_{sea}$ in $c\overline{c}$ & - & $1.5^*$ &  0.2 \\
em effects in $B_s$ or $B_c$ & 0.1 & $1.0^*$ & 1.0 \\
em effects in $D_s$  & - & - & 0.7 \\
finite volume & 0.0 & 0.0 & 0.0 \\
\hline
Total (MeV) & 11 & 9.5 & 19 \\
\hline
\hline
\end{tabular}
\caption{\label{tab:errors} Full error budget 
for $B_s$ and $B_c$ meson masses in MeV. See the text for 
a fuller description of each error. The total errors
are obtained by adding the individual errors in 
quadrature except in the case of the starred 
errors. They are correlated and this must be taken into 
account as described in the text
before being squared and accumulated into the square of the total. }
\end{table}

Additional systematic errors that must be added in 
to the error budget are those from relativistic 
corrections that are not included in our NRQCD 
action. These errors affect results at all 
lattice spacings equally and so cannot be 
estimated from our results as we 
have done for the discretisation errors. 
We must consider the effect of relativistic 
corrections on both the $B_s$ mass and on the 
$\Upsilon$ and $\eta_b$ masses because they both 
appear in $\Delta_{B_s}$. In fact we expect relativistic 
corrections to have a bigger effect in bottomonium 
than on the $B_s$. Our NRQCD action is correct 
through $\cal{O}$$(v_b^4)$ for bottomonium and 
so the largest missing terms are at $\alpha_sv_b^4$ 
and $v_b^6$. We expect the typical 
energy shift of a spin-independent 
$v_b^4$ term to be $\approx$ 50 MeV 
($0.1 \times$ 500 MeV), so $\alpha_sv_b^4$ corrections 
could give rise to 15 MeV shifts in $M_{b\overline{b}}$.   
Similarly a spin-independent $v_b^6$ correction could give rise 
to an energy shift of $\approx 5$ MeV. Adding these two 
in quadrature and dividing by 2 gives an estimate 
of the systematic error in $M_{B_s}$ from relativistic 
corrections to $M_{b\overline{b}}$ of 8 MeV.  
For $B_s$, the appropriate power counting for 
relativistic corrections is in $v_b \approx \Lambda/m_b$. 
Our NRQCD action already includes high order terms 
in $\Lambda/m_b$ at tree level and so there are no 
significant tree level errors for the $B_s$. 
The leading error is at $\alpha_s v_b$ in missing 
radiative corrections to $c_4$, the coefficient 
that multiplies the $\sigma \cdot B$ term giving 
rise to the hyperfine splitting. As we will discuss 
in section~\ref{se:hyp} we have plenty of evidence 
that errors coming from this term are small, at most 
10\% of the hyperfine splitting itself. Since this
error would vanish for the spin-average of the masses of the $B_s$ 
and the $B_s^*$, which is 3/4 of the hyperfine splitting 
above the $B_s$ mass, we take the error in the 
$B_s$ mass to be 3/4 of the error in the hyperfine 
splitting, 3.5 MeV.  

Errors from the uncertainties in $M_{\eta_s}$ and 
$M_{b\overline{b}}$ that we use to tune the $s$ and 
$b$ quark masses can be estimated from the slopes 
in Figures~\ref{fig:mbstunes} and~\ref{fig:mbstuneb}. 
The 4 MeV uncertainty in $M_{\eta_s}$ feeds into a 
1 MeV uncertainty in $\Delta_{B_s}$ and therefore 
$M_{B_s}$. The uncertainty in $\Delta_{B_s}$ from 
the 5 MeV uncertainty (simply adding the statistical 
and systematic errors) in $M_{b\overline{b}}$ is 
very small - less than 0.1 MeV because of the 
cancellation of the $b$ quark mass inside $\Delta_{B_s}$. 
However, the uncertainty reappears when we reconstruct 
$M_{B_s}$ by adding $M_{b\overline{b}}/2$ to $\Delta_{B_s}$.  
This then gives a sizeable 2.5 MeV error in $M_{B_s}$. 

Finite volume errors are expected from chiral 
perturbation theory to be negligible 
for the masses of mesons containing heavy quarks on volumes 
exceeding $(2.4 {\rm fm})^3$, that we are using here. 

The full error budget is given in Table~\ref{tab:errors}. 
The systematic errors discussed above, added in quadrature, 
give 9 MeV, dominating the 6 MeV errors coming from 
the statistical errors of the data and its lattice spacing 
and sea quark mass dependence. The final result is 
then $M_{B_s}$ = 5.363(6)(9) GeV.
Figure~\ref{fig:mbsasq} shows 
a dark shaded band for the first error and a lighter 
shaded band to encompass the full error, adding 6 MeV and 
9 MeV in quadrature to give 11 MeV. To reduce the 
full error will require improvements to the NRQCD action, 
currently underway. The experimental result for the $B_s$ mass is 
5.3663(6) GeV.  

\subsection{$B_c$ mass}
For the $B_c$ meson mass we could use exactly the 
same procedure as for the $B_s$. However, there 
is a better method, in which we subtract in addition 
the mass of a charmonium reference state, the $\eta_c$, to reduce the energy 
difference calculated on the lattice to a very small 
value: 
\begin{eqnarray}
M_{B_c} &=& \left(E_{B_c} -\frac{1}{2}(E_{b\overline{b}}
+M_{\eta_c})\right)_{\rm latt}a^{-1} \nonumber \\ 
&+& \frac{1}{2}\left(M_{b\overline{b},{\rm phys}} + M_{\eta_c,{\rm phys}}\right).
\label{EXPII}
\end{eqnarray}
We call this the ``heavy-heavy'' (hh) subtraction method. 
Here $M_{\eta_c}$ is the value of the $\eta_c$ mass 
calculated on the lattice and $M_{\eta_c,{\rm phys}}$ is its 
value from experiment 
appropriately adjusted for the lattice QCD world, 
as described earlier.   
For the HISQ quarks that we use for $c$ the energy 
obtained from fits to charmonium correlators at 
zero momentum is the charmonium mass, so there is 
no issue with the zero of energy. We simply use 
the additional charmonium subtraction here to reduce 
errors from the uncertainty in the lattice spacing. 

We will compare results of this to a second method:
\begin{eqnarray}
\label{EXPIII}
M_{B_c} &=& \left(E_{B_c} - (E_{B_s}+M_{D_s})\right)_{\rm latt}a^{-1}  \nonumber \\
&+& \left(M_{B_s, {\rm phys}} + M_{D_s, {\rm phys}}\right),
\end{eqnarray}
that we call the ``heavy-strange'' (hs) subtraction method. 
Here we are using the $B_s$ meson to cancel the NRQCD 
shift of the zero of energy in the $B_c$. Again the 
subtraction of the $D_s$ meson mass, calculated with 
HISQ $c$ and $s$ quarks is simply to reduce lattice 
spacing errors from the mass difference. 

\begin{figure}
\begin{center}
\includegraphics[width=80mm]{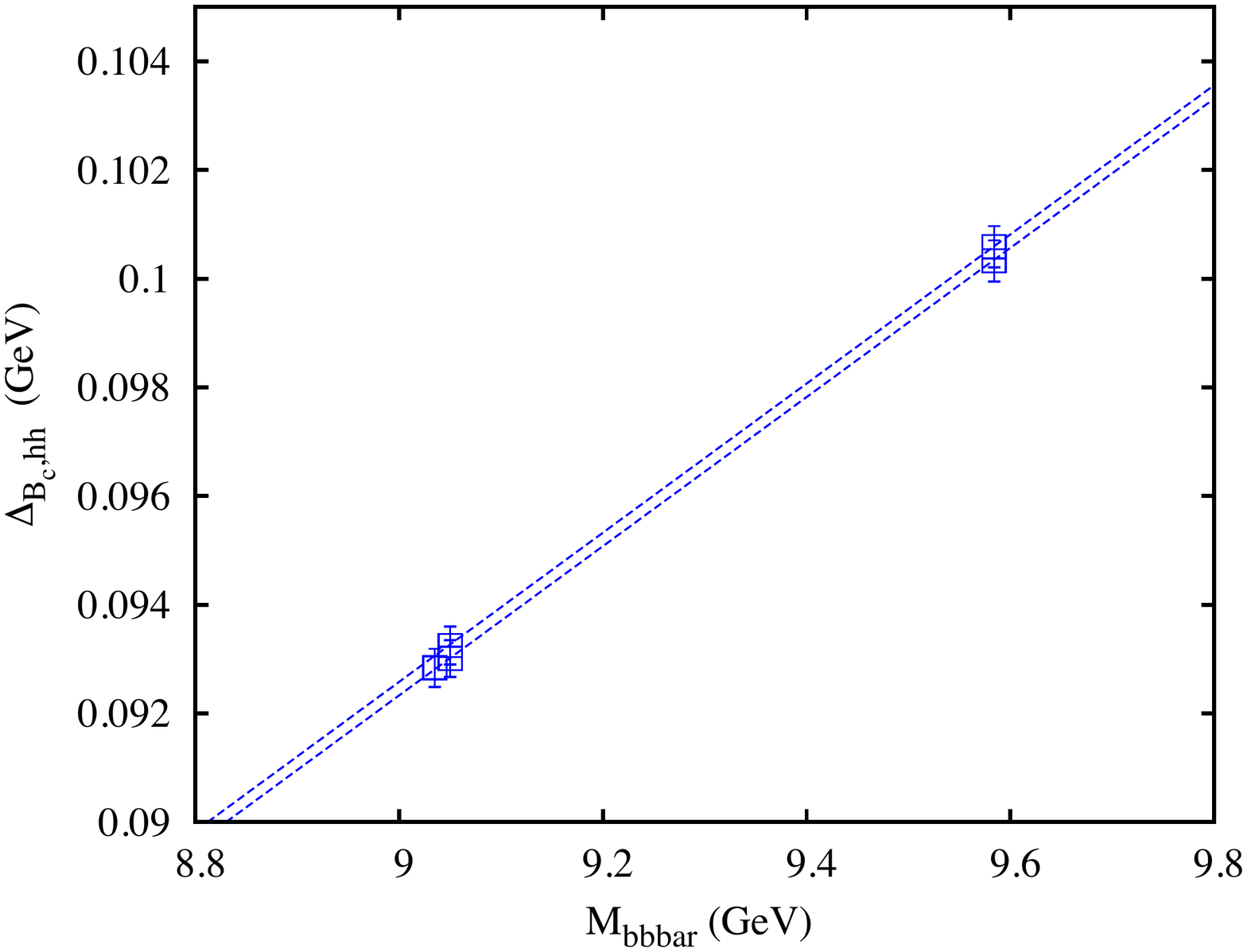}
\end{center}
\caption{Results from set 1 for the mass of the $B_c$ meson 
(specifically the difference between that mass and one 
half of the spin-averaged mass of $\Upsilon$ and $\eta_b$ 
added to the mass of the $\eta_c$ meson) 
as a function of the spin-averaged kinetic mass of the 
$\Upsilon$ and $\eta_b$, acting as a proxy for the $b$ quark mass. 
The errors are statistical only, since lattice spacing errors 
affect all the points together. 
The lines are fits to the results allowing 
linear terms in $M_{\eta_c}$ and $M_{b\overline{b}}$. Here the lines 
join points for a fixed $c$ quark mass. See Figure~\ref{fig:mbchhtunec}
for the equivalent as a function of $M_{\eta_c}$. 
The third point at lower left gives results for $c_i \ne 1$; 
it is not included in the fit. 
}
\label{fig:mbchhtuneb}
\end{figure}

\begin{figure}
\begin{center}
\includegraphics[width=80mm]{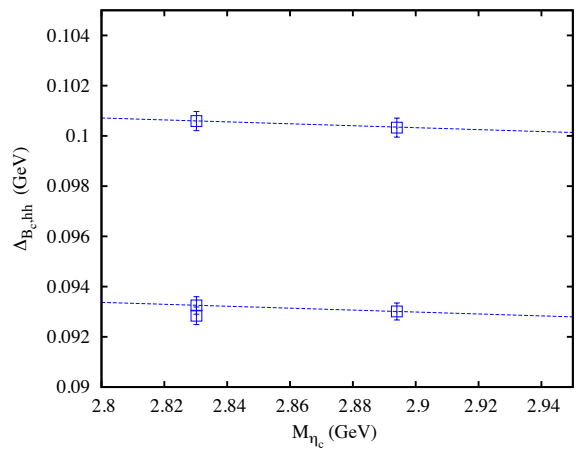}
\end{center}
\caption{Results from set 1 for the mass of the $B_c$ meson 
(specifically the difference between that mass and one 
half of the spin-averaged mass of $\Upsilon$ and $\eta_b$ 
added to the mass of the $\eta_c$ meson) 
as a function of the mass of the $\eta_c$ meson,
acting as a proxy for the $c$ quark mass. 
The errors are statistical only, since lattice spacing errors 
affect all the points together. 
The lines are fits to the results allowing 
linear terms in $M_{\eta_c}$ and $M_{b\overline{b}}$. Here the lines 
join points for a fixed $b$ quark mass. See Figure~\ref{fig:mbchhtuneb}
for the equivalent as a function of $M_{b\overline{b}}$. 
The third point at lower left gives results for $c_i \ne 1$; 
it is not included in the fit. 
}
\label{fig:mbchhtunec}
\end{figure}

We first discuss results from the hh method. The $B_c$ 
energies and $M_{\eta_c}$ masses are  
given in Table~\ref{tab:bcresults} and the $b\overline{b}$ 
energies and masses, already used in the determination of 
the $B_s$ mass, are given in Table~\ref{tab:bsresults}. 
As before, we have to tune quark masses on each ensemble 
to their correct value. We show in Figure~\ref{fig:mbchhtuneb} 
how the splitting 
\begin{equation}
\Delta_{B_c,hh} 
= \left(E_{B_c} -\frac{1}{2}(E_{b\overline{b}}
+M_{\eta_c})\right)_{\rm latt}a^{-1} 
\end{equation}
depends on $M_{b\overline{b}}$ from our results on 
set 1 at two values of $am_b$ and two values of $am_c$. We see that the slope 
is very small, 0.014, because very little $b$ quark 
mass dependence is left after cancellation in this
mass difference. An estimate can be derived for 
the expected slope by comparing results for the 
$b$ quark mass set to the value of the $c$ quark 
mass (when $\Delta_{B_c,hh}$ becomes -3/8 times
the charmonium hyperfine splitting). This gives 
a slope of 0.016 over a much wider range. 
Figure~\ref{fig:mbchhtuneb} also shows 
the value of the mass difference for the case where
we use an NRQCD action with $c_1$, $c_5$ and $c_6$
set to the values including $\cal{O}$$(\alpha_s)$ 
radiative corrections appropriate 
for set 1. 
We see that this makes negligible difference. 

\begin{table*}
\begin{ruledtabular}
\begin{tabular}{llllllll}
Set & $am_{b}$ & $am_{c}$ & $aM_{\eta_c}$ 
 & $aE_{B_c}$  & $a\Delta^{hyp}_c$ & $am_s$ & $aM_{D_s}$ \\
\hline
1 & 3.4 & 0.85 & 2.27031(16) & 1.34917(27) & 0.0324(2) & 0.066 & 1.5138(7) \\
  &     & 0.85 &         &  &  & 0.080 & 1.5295(10) \\
  & 3.4 & 0.88 & 2.32148(14) &  1.37456(27) & 0.0325(2) & 0.066 & 1.5441(14) \\
  &     & 0.88 &   &  &  & 0.080 & 1.5587(10) \\
  & 3.6 & 0.85 &             & 1.35415(29) & 0.0309(3) &  &  \\
  & 3.6 & 0.88 &             & 1.37593(29) & 0.0311(3) &  &  \\
  & 3.4($c_i \neq 1$) & 0.85 &           & 1.34987(27) & 0.0323(2) & & \\
2 & 3.4 & 0.85 & 2.26964(17) & 1.34834(34) & 0.0326(3) & 0.066 & 1.5140(8) \\
\hline
3 & 2.8 & 0.65 & 1.84949(11) & 1.11727(13) & 0.0268(2) & 0.0537 & 1.2260(5) \\
4 & 2.8 & 0.66 & 1.87142(12) & 1.12783(25) & 0.0271(4) & 0.05465 & 1.2406(5) \\
\hline
5 & 1.95 & 0.43 & 1.31691(7) & 0.81861(12) & 0.0210(2) & 0.0366 & 0.8709(3) \\
\end{tabular}
\end{ruledtabular}
\caption{\label{tab:bcresults} Results for energies and masses 
needed for the determination of the mass of the $B_c$ meson.  
All the results are in units of the lattice spacing. 
For each valence $b$ quark mass 
the kinetic mass and energy for the spin-average of $\Upsilon$ and 
$\eta_b$ states is given in Table~\ref{tab:bsresults} as are 
the $\eta_s$ meson masses for each $s$ quark mass and the corresponding  
$B_s$ meson energies. 
In column 4 we give the $\eta_c$ meson mass 
corresponding to each value of $am_c$, and in the final column 
we give the corresponding $D_s$ meson mass. 
Note that the values 
for $M_{\eta_c}$ and $M_{D_s}$ are different from those 
reported in~\cite{newfds} because here 
we are using a nonperturbatively determined Naik coefficient 
as discussed in the text.  
In column 5, we give the fitted energy of the $B_c$ meson 
(i.e. $E_0$ from fits to the form given in equation~\ref{eq:fitform}). 
In column 6 we give 
the hyperfine splitting, $\Delta^{hyp}_c = E(B_c^*)-E(B_c)$, 
discussed in section~\ref{se:hyp}.
This column is 
largely from~\cite{Gregory:2009hq} but includes some additional 
values on set 1 that we use for studying systematic errors. 
}
\end{table*}

Figure~\ref{fig:mbchhtunec} shows the results as 
a function of $M_{\eta_c}$. The slope here is 
very small but in the opposite direction to 
that for the dependence on $M_{b\overline{b}}$. 
The value of the slope is -0.004. 
Based on the arguments above we would expect a 
slope of opposite sign but about 60\% that
of the $b$-quark mass dependence. 
However, as stated above, this estimate is made over a much 
larger range than the Figure. 

We can use the slope against $M_{b\overline{b}}$ and 
against $M_{\eta_c}$ to correct 
for the slight mistunings of the $b$ quark and the $c$ quark 
that we have on 
some ensembles. Even though the shifts from $M_{b\overline{b}}$ 
and $M_{\eta_c}$ 
dependence are very small they are not negligible. This is because 
$\Delta_{B_c, hh}$ itself is very small and also because it 
is very precise, since all of the energies involved have 
tiny statistical errors. We take a 50\% error on $b$ quark 
mistuning but a 200\% error from $c$ quark mistuning to allow 
for the fact that we may be underestimating the slope with 
$c$ quark mass because of discretisation errors in the HISQ 
action for $c$ on the very coarse lattices. 

\begin{table}[ht]
\begin{tabular}{ccc}
\hline
\hline
Set & $\Delta_{B_c,hh}$ (GeV) & -$\Delta_{B_c,hs}$ (GeV)\\
\hline
1 & 0.0980(4)(12)(8) & 1.034(2)(16)(4) \\
2 & 0.0974(4)(35)(9) & 1.035(2)(28)(4) \\
\hline
3 & 0.0782(2)(5)(6) & 1.044(3)(9)(4) \\
4 & 0.0788(3)(9)(7) & 1.046(3)(10)(5)\\
\hline
5 & 0.0652(3)(14)(5) & 1.054(2)(2)(4)\\
\hline
\hline
\end{tabular}
\caption{\label{tab:tunebc} Results for $\Delta_{B_c}$, 
the mass difference between the $B_c$ meson 
and a particular reference mass,
on different ensembles 
after tuning to the correct valence $b$, $c$  and, where 
appropriate, $s$ quark masses.
The 3 errors listed are statistical, tuning and from the 
uncertainty in the lattice spacing. 
Column 2 gives results from the hh method and column 
3 from the hs method, as described
in the text.}
\end{table}

\begin{figure}
\begin{center}
\includegraphics[width=80mm]{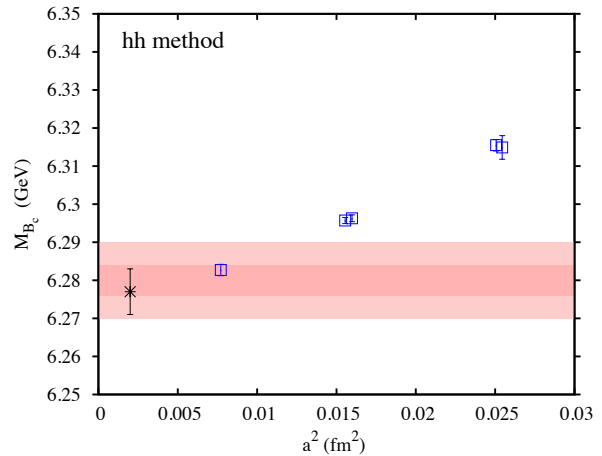}
\end{center}
\caption{Results for the mass of the $B_c$ meson 
tuned to the correct $c$ and $b$ quark masses on 
each ensemble obtained from the `hh method' 
and plotted against the square of 
the lattice spacing. The errors on the points 
include statistical and tuning errors and the 
uncertainty in the lattice spacing. 
The dark shaded band is our physical result, allowing 
for sea quark mass and lattice spacing 
dependence as described in 
the text, and including electromagnetic effects. 
The width of the lighter shaded  
band reflects our full error as given in the error 
budget, Table~\ref{tab:errors}. The black star is the 
experimental result~\cite{pdg}, offset from $a^2=0$ 
for clarity.  
}
\label{fig:mbcasq}
\end{figure}

Table~\ref{tab:tunebc} gives tuned values for $\Delta_{B_c,hh}$ 
on each ensemble, along with three errors; that from statistics, 
from tuning and from the uncertainty in the lattice spacing. 
These latter two errors dominate. 
As before, variation in the value of the lattice spacing means 
that masses must be retuned. Here this has the effect of producing 
a net change equal to the naive lattice spacing error. 
Figure~\ref{fig:mbcasq} plots these results against the 
square of the lattice spacing, after reconstructing the 
$B_c$ mass by adding back in $(M_{b\overline{b},{\rm phys}} + M_{\eta_c,{\rm phys}})/2$ = 
6.2175 GeV. 

Clear lattice spacing dependence is visible in Figure~\ref{fig:mbcasq} 
but there is no sign of sea quark mass dependence. 
As for the case of $\Delta_{B_s}$ we fit the results for $\Delta_{B_c,hh}$ 
as a function of 
lattice spacing and sea quark mass to extract a physical result. 
The fit form is essentially the same as for $\Delta_{B_s}$. However, 
because $\Delta_{B_c,hh}$ is such a small quantity it cannot 
set the scale for the discretisation and sea quark mass effects. 
So, instead of allowing a function of $a$, $\delta x_l$ and $\delta x_s$ 
to multiply $\Delta_{B_c,hh}$ we add such a function with a multiplicative 
factor of 0.4 GeV, representing a typical scale for QCD binding energies. 
We include more terms for discretisation errors than in the 
$B_s$ case and set their 
scale by $m_c$, rather than $\Lambda$, because in this case discretisation 
errors will come largely from the HISQ action for the $c$ quark. 
\begin{eqnarray}
\label{eq:fitbcextrap}
\Delta_{B_c,hh}(a,\delta x_l, \delta x_s) &=& \Delta_{B_c,hh,{\rm phys}} + \\ \nonumber 
0.4\big[ \sum_{j=1}^4 c_j(m_c a)^{2j}(&1& + c_{jb}\delta x_m + c_{jbb}(\delta x_m)^2) \\ \nonumber 
+ 2b_l\delta x_l(&1&+c_l(m_c a)^2+c_{ll}(m_c a)^4) \\ \nonumber
+ 2b_s\delta x_s(&1&+c_s(m_c a)^2+c_{ss}(m_c a)^4) \\ \nonumber
+  4b_{ll}(\delta x_l)^2 &+& 2b_{ls}\delta x_l\delta x_s + b_{ss}(\delta x_s)^2\big] .
\end{eqnarray} 
We take the prior on $\Delta_{B_c,hh,{\rm phys}}$ to be 0.05(5). The priors 
on the sea quark mass dependence, $b_l$ and $b_s$, are 
taken to be 0.00(7), and on the parameters corresponding to the 
quadratic sea mass dependence 0.000(13), as before.  
We take the scale of the physical $a$-dependence to be the scale 
of $m_c \approx$ 1 GeV, as discussed above. 
The coefficients of the quadratic $a$ dependence, $c_1$, $c_l$ and $c_s$, 
should be $\cal{O}$$(\alpha_s)$ so we take priors of 0.0(5). For $c_2$, 
and the $am_b$ dependence of the discretisation errors, $c_{jb}$ and $c_{jbb}$, we take 
a very conservative prior of 0(1). 

The result for $\Delta_{B_c,hh,{\rm phys}}$ is 0.0616(42) GeV, 
giving a $B_c$ mass of 6.279(4) GeV. The fit is robust 
under changes of the prior values. For example, we tried 
the following changes:
\begin{itemize}
\item taking the multiplier of $a$- and $m_{sea}$-dependence 
to be 0.8 instead of 0.4; 
\item taking the priors for $a$-dependence to have width 2 
instead of 1; 
\item taking the priors on sea quark mass dependence to 
be 0.5 rather than 0.2.
\end{itemize}
None of these changed the result by more than 1 MeV. 

Our result is for a world without electromagnetism or charm 
quarks in the sea. The effects 
of electromagnetism on the $B_c$ can be estimated from a potential model 
in the same way as we have done for bottomonium and charmonium. 
The quark and antiquark in the $B_c$ have the same sign of electric 
charge, however, and so the effect now is repulsive rather than 
attractive. We estimate that the effect of switching on electromagnetism 
is to shift our $B_c$ mass upwards by 2 MeV. The effects of $c$ quarks 
in the sea can be estimated following the discussion in 
subsection~\ref{subse:NRQCD} as approximately 1/60 of the hyperfine 
splitting in the $B_c$ system, or 1 MeV. This effect is attractive 
and so will counteract the effect of electromagnetism. We take 
the net shift in the $B_c$ mass as 1(1) MeV, moving our result 
 to 6.280(4) GeV. This
is the value given by the dark shaded band in Figure~\ref{fig:mbcasq}.  

The complete error budget is given in Table~\ref{tab:errors}. 
Here we have split up the 4 MeV error from the fit into 
its component parts as discussed in the $B_s$ case and 
added sources of systematic error. 
Errors from missing relativistic corrections to the NRQCD action 
are similar to those for the $B_s$ case. The leading missing spin-independent 
terms in the NRQCD action are $\cal{O}$$(\alpha_sv_b^4)$ and 
$\cal{O}$$(v_b^6)$. We must estimate the effect of these terms on 
both $M_{b\overline{b}}$ and $M_{B_c}$. 
$v_b^2$ is about half the 
size in the $B_c$ compared to bottomonium. This means that 
there is some cancellation of the $\alpha_sv_b^4$ errors in 
$\Delta_{B_c,hh}$, since we estimated $\alpha_sv_b^4$ errors previously 
as $\alpha_sv_b^2$(500 MeV). Independent confirmation that these 
terms have a small net effect comes from the calculations 
that we have done here for the case where $c_1$, $c_5$ and $c_6$ 
take the values that include the $\cal{O}$$(\alpha_s)$ radiative 
corrections, see Table~\ref{tab:bcresults} and Figures~\ref{fig:mbchhtuneb} 
and~\ref{fig:mbchhtunec}. 
There will be little cancellation of the 
$v_b^6$ errors, however, since $v_b^6$ is much smaller in the 
$B_c$ than in bottomonium. We take the systematic error from 
spin-independent terms then to be 4 MeV for $\alpha_s v_b^4$ (i.e. 
half that for $B_s$) and 5 MeV for $v_b^6$, added in quadrature 
to give 6 MeV. The leading spin-dependent error is from missing 
radiative corrections to the $\sigma.B$ term. This affects 
the $B_c$ only because of the spin-averaging of $M_{b\overline{b}}$. 
As for the $B_s$ we take 3/4 of a possible 10\% error in the 
hyperfine splitting, estimated in~\cite{Gregory:2009hq} at 53 MeV, 
i.e. 4 MeV.      

Errors from the uncertainty in $M_{b\overline{b}}$ and 
$M_{\eta_c}$ do not affect $\Delta_{B_c,hh}$ significantly 
but $M_{B_c}$ inherits an error of half their uncertainty 
when it is reconstructed from $\Delta_{B_c}$ and those masses. 
These two uncertainties are partly correlated, because they both 
contain estimates of electromagnetic and annihilation 
effects in the two very similar charmonium and bottomonium 
systems. The systematic errors from electromagnetism and 
$c$ quarks in the sea for the $B_c$ are also correlated with 
the errors for these effects from charmonium and bottomonium. 
These errors are marked with a star in Table~\ref{tab:errors} 
and we separate out the different components and add them 
linearly with appropriate signs before squaring and accumulating into the
total error. The error from these three components is 
then increased by their correlation from 3 MeV to 4.6 MeV. 
We estimate finite volume errors to be negligible 
for the $B_c$. Our total error, from adding statistical and 
systematic errors in quadrature, is 9.5 MeV, giving a 
mass for the $B_c$ meson from the hh method of 
6.280(10) GeV. The total error is plotted as the more 
lightly shaded band in Figure~\ref{fig:mbcasq}.  

\begin{figure}
\begin{center}
\includegraphics[width=80mm]{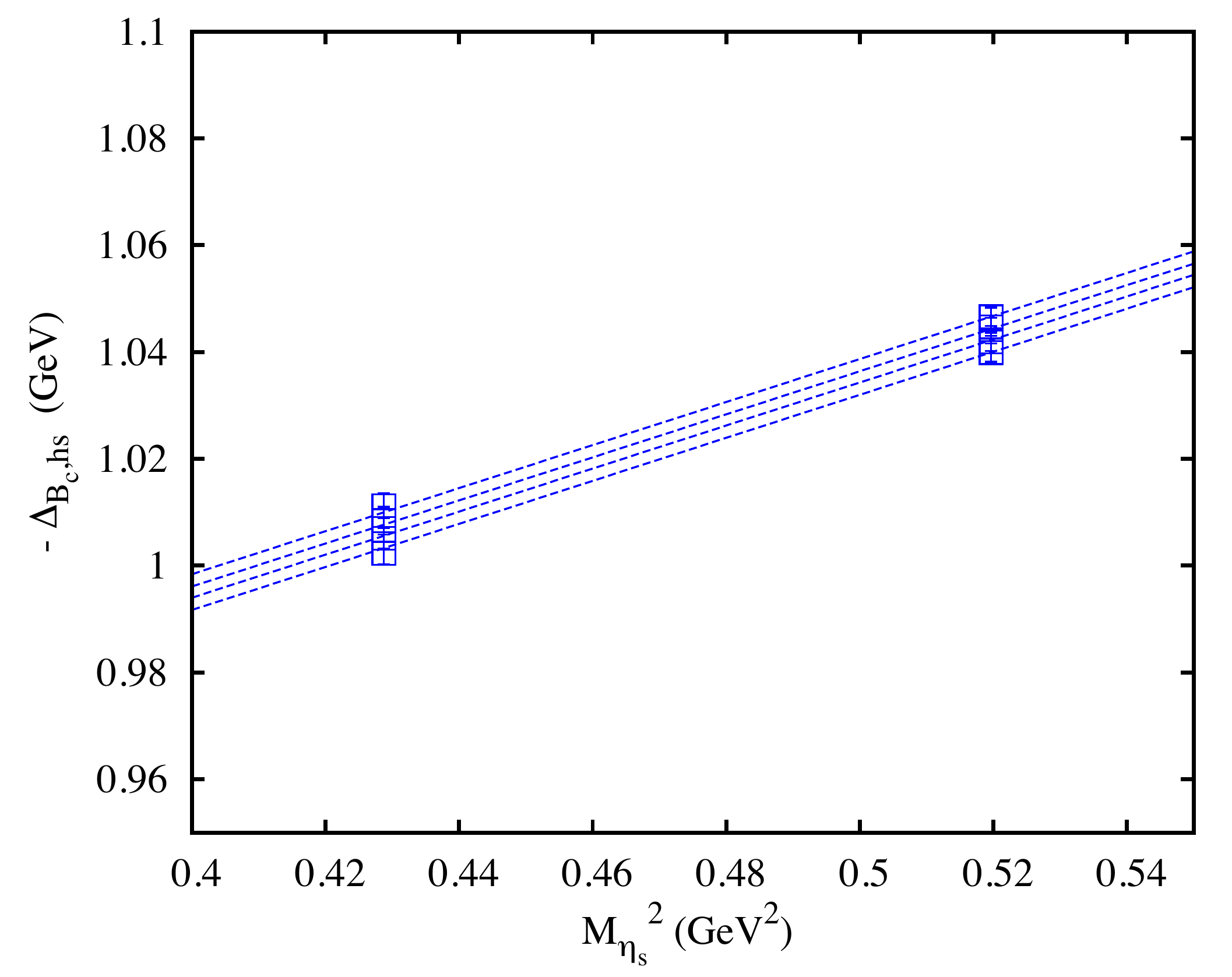}
\end{center}
\caption{Results from set 1 for the mass of the $B_c$ meson 
(specifically the difference between that mass and the 
masses of the $B_s$ and $D_s$ mesons)
as a function of the square 
of the $\eta_s$ meson mass, acting as a proxy for the 
strange quark mass. 
The errors are statistical only, since lattice spacing errors 
affect all the points together. 
The lines are fits to the results allowing 
linear terms in $M_{\eta_s}^2$, $M_{b\overline{b}}$ and $M_{\eta_c}$. 
Here the lines 
join points for fixed $b$ and $c$ quark masses. See Figure~\ref{fig:mbchstuneb}
for the equivalent as a function of $M_{b\overline{b}}$ and 
Figure~\ref{fig:mbchstunec} for the equivalent as a function of 
$M_{\eta_c}$. 
}
\label{fig:mbchstunes}
\end{figure}

\begin{figure}
\begin{center}
\includegraphics[width=80mm]{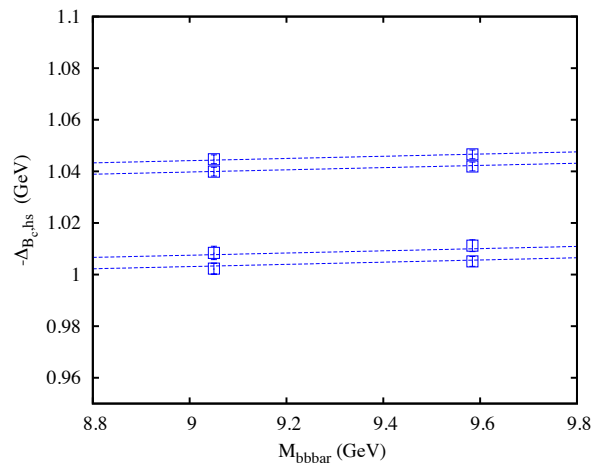}
\end{center}
\caption{Results from set 1 for the mass of the $B_c$ meson 
(specifically the difference between that mass and the 
masses of the $B_s$ and $D_s$ mesons)
as a function of the spin-average mass of the $\Upsilon$ 
and $\eta_b$, $M_{b\overline{b}}$,
acting as a proxy for the 
$b$ quark mass. 
The errors are statistical only, since lattice spacing errors 
affect all the points together. 
The lines are fits to the results allowing 
linear terms in $M_{\eta_s}^2$, $M_{b\overline{b}}$ and $M_{\eta_c}$. 
Here the lines 
join points for fixed $s$ and $c$ quark masses. See Figure~\ref{fig:mbchstunes}
for the equivalent as a function of $M_{\eta_s}^2$ and 
Figure~\ref{fig:mbchstunec} for the equivalent as a function of 
$M_{\eta_c}$. 
}
\label{fig:mbchstuneb}
\end{figure}

\begin{figure}
\begin{center}
\includegraphics[width=80mm]{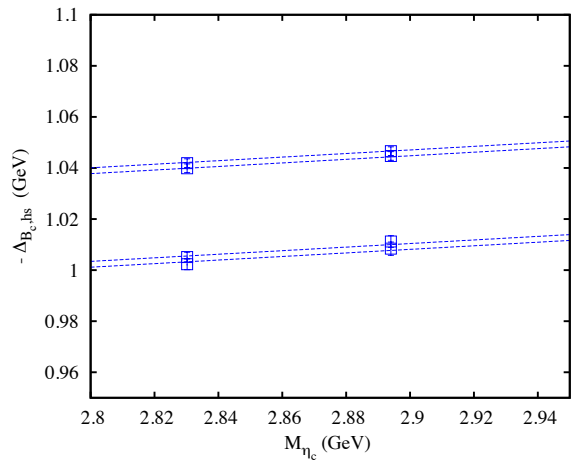}
\end{center}
\caption{Results from set 1 for the mass of the $B_c$ meson 
(specifically the difference between that mass and the 
masses of the $B_s$ and $D_s$ mesons)
as a function of the mass of the $\eta_c$ 
acting as a proxy for the 
$c$ quark mass. 
The errors are statistical only, since lattice spacing errors 
affect all the points together. 
The lines are fits to the results allowing 
linear terms in $M_{\eta_s}^2$, $M_{b\overline{b}}$ and $M_{\eta_c}$. 
Here the lines 
join points for fixed $s$ and $b$ quark masses. See Figure~\ref{fig:mbchstunes}
for the equivalent as a function of $M_{\eta_s}^2$ and 
Figure~\ref{fig:mbchstuneb} for the equivalent as a function of 
$M_{b\overline{b}}$. 
}
\label{fig:mbchstunec}
\end{figure}

The hs method has different systematic errors from the hh method 
and so provides a good cross-check. 
The raw results needed for this method are given in Table~\ref{tab:bcresults} 
and in Figures~\ref{fig:mbchstunes},~\ref{fig:mbchstuneb} and~\ref{fig:mbchstunec} we show results for the quantity $-\Delta_{B_c,hs}$ 
(because $\Delta_{B_c,hs}$ is negative) defined by:
\begin{equation}
\Delta_{B_c,hs} 
= \left(E_{B_c} -(E_{B_s}
+M_{D_s})\right)_{\rm latt}a^{-1}. 
\end{equation}
In the figures $-\Delta_{B_c,hs}$ results from set 1 are 
plotted against the different 
quark masses involved in the calculation, 
with $M_{\eta_s}^2$, $M_{b\overline{b}}$ and
$M_{\eta_c}$ acting as proxies for the $s$, $b$ and $c$ quark 
masses respectively (we have two different values for the 
masses of each quark). We see that there is fairly 
strong dependence on the $s$ quark mass but very little 
on the $b$ quark mass or the $c$ quark mass. 
The slope against $M_{\eta_s}^2$ is 0.41, which agrees well with 
that expected from experiment if we substitute light quarks 
for $s$ quarks in the formula for $\Delta_{B_c,hs}$ above. 
The slope against $M_{b\overline{b}}$ is 0.005 and against 
$M_{\eta_c}$ is 0.07. These are only in very rough agreement with 
the expectations of 0.02 and 0.04 respectively from 
comparing results over the much larger experimental 
range from $b$ to $c$~\cite{pdg}.  

As before we can use these results to compensate for 
mistuning of the quark masses. Again we take a 50\% error 
on tuning shifts for $b$ and $s$ but a 200\% errors on those 
for $c$ to allow for discretisation errors in our estimates 
of those effects. The tuned results 
are given in Table~\ref{tab:tunebc}. All of the statistical, 
tuning and lattice spacing errors are larger than 
those of the hh method. The tuning error dominates 
on the very coarse and coarse lattices, but on the fine 
ensemble it is comparable to the other errors. 
The 
lattice spacing error is reduced by a factor of two over the 
naive error by the retuning required when the lattice 
spacing changes. Lattice spacing dependence is small 
but visible in these results; no sea quark mass dependence 
is evident. 

\begin{figure}
\begin{center}
\includegraphics[width=80mm]{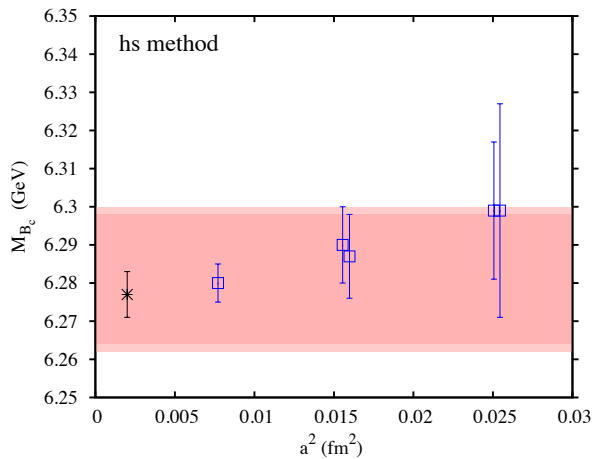}
\end{center}
\caption{Results for the mass of the $B_c$ meson 
tuned to the correct $c$ and $b$ quark masses on 
each ensemble obtained from the `hs method' 
and plotted against the square of 
the lattice spacing. The errors on the points 
include statistical and tuning errors and the 
uncertainty in the lattice spacing. 
The darker shaded band is our physical result, allowing 
for sea quark mass and lattice spacing 
dependence as described in 
the text, and including electromagnetic effects. 
The width of the lighter shaded 
band reflects our full error as given in the error 
budget, Table~\ref{tab:errors}. The black star is the 
experimental result~\cite{pdg}, offset from $a^2=0$ for 
clarity.  
}
\label{fig:mbchsasq}
\end{figure}

As for the $B_s$ and the $B_c$ hh method we fit 
the results for $\Delta_{B_c,hs}$ to a functional form allowing for 
lattice spacing dependence, including NRQCD 
effects, and sea quark mass dependence. 
The functional form is the same as that used
for the hh method, except that now 
we can include these dependences as a multiplicative 
factor since $\Delta_{B_c,hs}$ is not unusually 
small. We use: 
\begin{eqnarray}
\label{eq:fitbchsextrap}
\Delta_{B_c,hs}(a,\delta x_l, \delta x_s) &=& \Delta_{B_c,hs,{\rm phys}}\big[ 1 + \\ \nonumber 
 \sum_{j=1}^4 c_j(m_c a)^{2j}(&1& + c_{jb}\delta x_m + c_{jbb}(\delta x_m)^2) \\ \nonumber 
+ 2b_l\delta x_l(&1&+c_l(m_c a)^2+c_{ll}(m_c a)^4) \\ \nonumber
+ 2b_s\delta x_s(&1&+c_s(m_c a)^2+c_{ss}(m_c a)^4) \\ \nonumber
+  4b_{ll}(\delta x_l)^2 &+& 2b_{ls}\delta x_l\delta x_s + b_{ss}(\delta x_s)^2\big] .
\end{eqnarray} 
We take the same prior values as for the hh method except for 
the prior for $\Delta_{B_c,hs,{\rm phys}}$ which we take to be -1.0(2). 

We obtain the result $\Delta_{B_c,hs,{\rm phys}}$ = -1.054(17) GeV. 
This changes by less than 1 MeV on doubling the prior width 
for the lattice spacing dependence or the sea quark mass 
dependence. 
To reconstruct the $B_c$ mass from this we need to add 
the appropriate values for the $B_s$ and $D_s$ masses in 
a world without electromagnetism or $c$ quarks in the 
sea. As discussed earlier, electromagnetism has negligible 
effect on the $B_s$ mass. The $D_s$ mass is lower by 1.3(7) MeV, 
however, in a world without electromagnetism, from the 
phenomenological formula in equation~\ref{eq:em}. This gives a 
total for the appropriate sum of $M_{D_s} + M_{B_s}$ of 
7.334 GeV. Figure~\ref{fig:mbchsasq} shows our tuned results 
for $m_{B_c}$ as a function of lattice spacing. 
The dark shaded band is the result from the fit just described 
including the error obtained from it. We have shifted the 
$B_c$ mass obtained upwards by 1(1) MeV, to a central 
value of 6.281 GeV, as described 
earlier to allow for electromagnetic and charm-in-the sea 
effects that are not included in our calculation. The 
lighter shaded band gives the total error, from the 
error budget of Table~\ref{tab:errors}, the systematic error components 
of which we will now discuss. The first three entries in the 
Table are the split of the 17 MeV error obtained from 
the fit among its different components. 

Errors from missing relativistic corrections to the NRQCD action 
affect the $B_c$ energy and the reference $B_s$ energy. 
The leading missing spin-independent 
terms in the NRQCD action are $\cal{O}$$(\alpha_sv_b^4)$ and 
$\cal{O}$$(v_b^6)$. We earlier estimated the shift from 
$\cal{O}$$(\alpha_sv_b^4)$ terms on $M_{b\overline{b}}$ at 
15 MeV. For the $B_c$ we expect a systematic error of about 
half this value, so we take 7.5 MeV, since $v_b^2$ is roughly half as big. 
For $B_s$ $\alpha_sv_b^4$ terms have very little effect and 
neither meson will be sensitive to $v_b^6$ terms. 
Spin-dependent NRQCD errors come chiefly from missing radiative 
corrections to the $\sigma\cdot B$ term, but there will be 
cancellation here between the $B_c$ and the $B_s$ since both 
mesons will respond in the same way to a change in $c_4$. 
We therefore take a systematic error of 1 MeV rounding up the 
difference between the 4 MeV systematic previously allowed 
for this for the 
$B_c$ and the 3.5 MeV systematic for the $B_s$. 

Systematic errors from uncertainties in the physical values 
of $M_{\eta_s}$, $M_{\eta_c}$ and $M_{b\overline{b}}$ which 
affect the quark mass tuning can be estimated from the 
dependence of $\Delta_{B_c,hs}$ on these quantities discussed 
earlier. The error from the uncertainty in $M_{\eta_s}$ 
is sizeable at 2.3 MeV; the others are very small. 
We must also allow for systematic errors from uncertainties 
from electromagnetism and charm-in-the-sea for the 
reference masses of the $D_s$ (0.7 MeV is half the shift 
applied in that case) and the $B_s$ (negligible) as 
well as for the $B_c$ itself (1 MeV as discussed above). 

This gives a total error of 19 MeV, dominated by the 
statistical and tuning errors of the raw data. 
Our final result for the $B_c$ mass from the 
hs method is then 6.281(19) GeV. This is plotted as the more 
lightly shaded band in Figure~\ref{fig:mbchsasq}.  
The agreement between the hs and hh methods is very good, 
although their systematic and statistical errors are 
very different, with the hh method being significantly more accurate. 
The agreement is in fact not surprising when we consider that the 
$B_s$ mass determined in section~\ref{subse:bs} agreed well 
with experiment. The $B_c$ hs method replaces $M_{b\overline{b}}$ 
with $M_{B_s}$ and $M_{\eta_c}$ with $M_{D_s}$ so if 
the $B_s$ and $D_s$ masses are known to agree with experiment 
given masses tuned from $M_{b\overline{b}}$ and $M_{\eta_c}$ 
then the $B_c$ from hs and hh will agree. 
However, the fact that they were derived completely independently 
is a good consistency check of the method and of our error 
estimates. 

\subsection{$B$ mass} 
\label{se:bmass}

The correlators for the $B$ meson are noisier than those for 
the $B_s$, as will be clear from the discussion of the 
signal/noise earlier. This means that the $B$ meson mass is 
the least well determined of all our masses. 
The best way 
then to pin down the $B$ mass is to consider the mass 
difference between the $B_s$ and the $B$. NRQCD systematic 
errors should entirely cancel in such a difference. However, 
because the difference is a small number we have a fairly sizeable 
statistical error even when we fit both correlators together, 
as described earlier, and extract $E_{B_s}-E_B$ directly 
from the fit. 

\begin{table*}[ht]
\begin{tabular}{llllccc}
\hline
\hline
Set & $am_b$ & $am_s$ & $am_l$ &  $aM_{\pi}$ & $aE(B_s)-aE(B_l)$ & $a\Delta^{hyp}_l$ \\
\hline
1 & 3.4 &  0.066 & 0.0132 & 0.2408(6) & 0.0553(62) & 0.0318(78)  \\
 & 3.4 & 0.080 & 0.0132 &  & 0.0683(62) &   \\
2 & 3.4 & 0.066 & 0.0264 & 0.3348(6)  & 0.0369(29) & 0.0374(55)  \\
\hline
3 & 2.8 & 0.0537 & 0.0067 &  0.1567(4) & 0.0446(46)  & 0.0306(54)  \\
4 & 2.8 & 0.05465 & 0.01365 & 0.2222(5) & 0.0336(41)   & 0.0245(68)  \\

\hline
5 & 1.95 & 0.0366 & 0.00705 & 0.1377(4)  & 0.0245(30)   & 0.0177(35)  \\
\hline
\hline
\end{tabular}
\caption{\label{tab:bresults} Results for energies and masses needed 
for the determination of the mass of the $B$ meson. 
Column 6 gives the energy difference between $B_s$ and $B_l$ mesons, 
for different valence $b$, $s$ and $l$ quark masses given in columns 2, 3 and 4. 
Column 5 gives the corresponding $\pi$ meson mass, taken from~\cite{kendall}. 
Column 7 gives the hyperfine splitting for the $B_l$ meson, from~\cite{Gregory:2009hq}, 
$\Delta^{hyp}_l = E(B_l^*)-E(B_l)$. }
\end{table*}

Table~\ref{tab:bresults} gives values for the energy difference between 
$B_s$ and $B$ extracted from our fits on each ensemble. Statistical errors 
are 10-15\% of the splitting. However, this amounts to less than 
10 MeV in terms of the absolute mass, so still provides a good 
test against experiment for $M_B$.   

\begin{figure}
\begin{center}
\includegraphics[width=80mm]{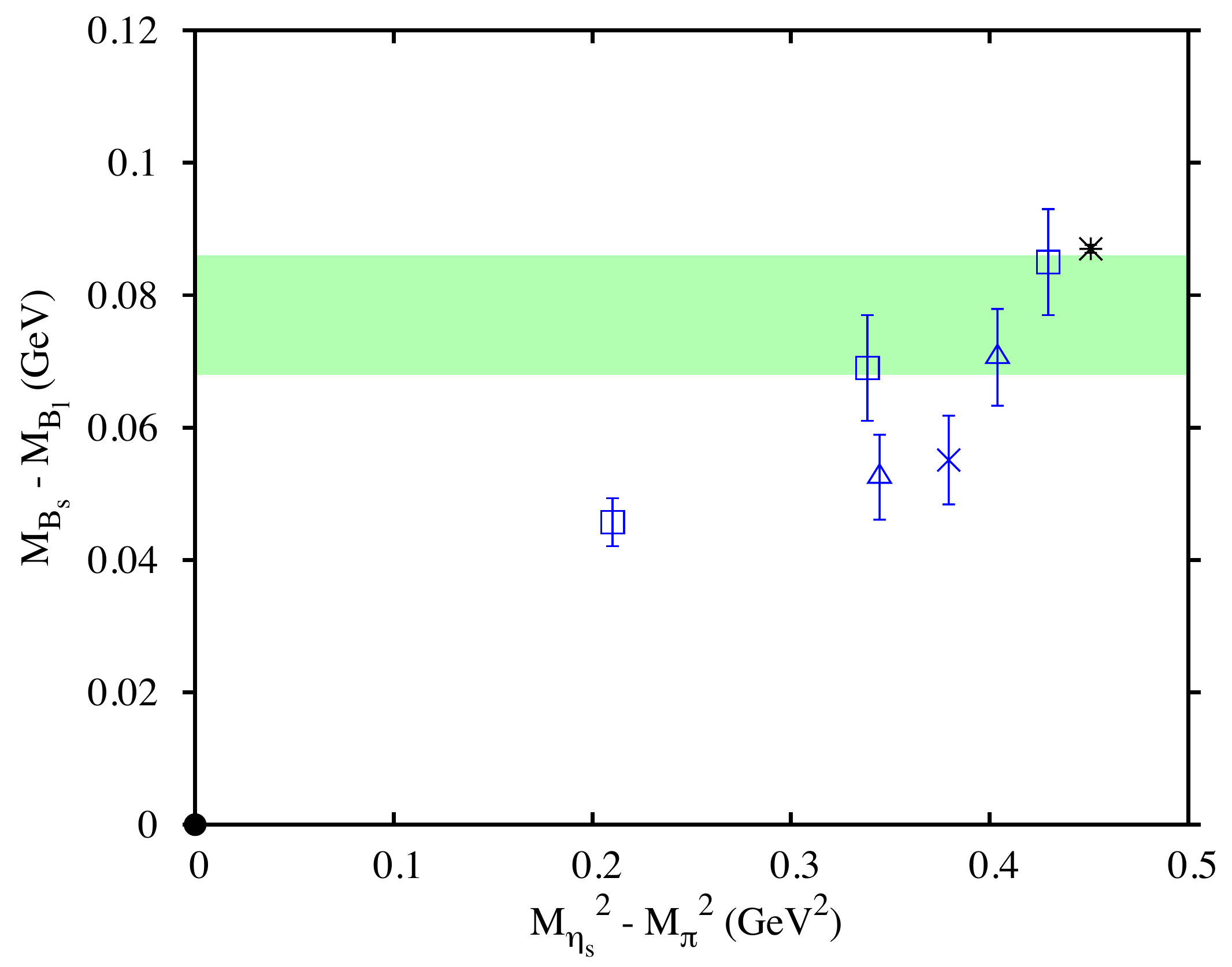}
\end{center}
\caption{Results for the mass difference between 
the $B_s$ and $B$ mesons as a function of 
$M_{\eta_s}^2-M_{\pi}^2$ acting as a proxy 
for $m_s-m_l$. Open squares are results on very 
coarse ensembles, triangles on coarse ensembles 
and the cross is from the fine ensemble. 
The green band represents the result of our fit, 
adjusted for electromagnetic effects 
as described in the text.  The black star is 
the experimental result for $M_{B_s}-(M_{B^0}+M_{B^\pm})/2$. 
}
\label{fig:bsbplot}
\end{figure}

We plot the results for $M_{B_s}-M_{B}$ (= $a^{-1}(aE_{B_s}-aE_{B})$)
against $M_{\eta_s}^2-M_{\pi}^2$ which is a useful physical proxy 
for $m_s-m_l$ in Figure~\ref{fig:bsbplot}. 
We expect this mass difference to be largely linear 
in $m_s-m_l$ and our results are consistent with this. Given the 
statistical errors, we fit a relatively simple form to this 
difference:  
\begin{eqnarray}
\label{eq:fitformb}
(M_{B_s}-M_B)(a,\delta x_l, \delta x_s) =  \nonumber \\ 
\sum_{j=1}^3 a_jf_j(M_{\eta_s}, M_{\pi})(1 + c_{j1}(\Lambda a)^2 + c_{j2}(\Lambda a)^4) \nonumber \\  
+ f_1(M_{\eta_s}, M_{\pi})(2b_l\delta x_l(1+c_l(\Lambda a)^2)) \nonumber \\ 
+ 2b_s\delta x_s(1+c_s(\Lambda a)^2).  
\end{eqnarray} 
Here the functions $f_j$ are simple ones that respect the fact that 
$M_{B_s}-M_B$ vanishes by definition when the light quark mass 
is equal to the strange quark mass. So $f_1 = M_{\eta_s}^2-M_{\pi}^2$, 
$f_2 = M_{\eta_s}^4 - M_{\pi}^4$ and $f_3 = M_{\eta_s}^2\log M_{\eta_s^2} - M_{\pi}^2 \log M_{\pi}^2$. We allow these terms to have lattice spacing dependence
with a scale set by $\Lambda$ = 400 MeV. We also allow sea quark mass 
dependence in the terms multiplying $f_1$. The coefficients $a_j$ are 
given priors of 0.0(5) (we expect a slope of 0.2 if the dependence 
on $M_{\eta_s}^2 - M_{\pi}^2$ were purely linear). For the 
$c_{j2}$ $a^4$-dependence coefficients we take 0.0(1.0) and for 
$c_{j1}$ we take 0.0(5) since $a^2$ terms should be suppressed 
by an additional power of $\alpha_s$. For the sea quark mass dependent 
coefficients $b_l$ and $b_s$ we take priors of 0.00(7) as discussed 
earlier. 

The physical value for $M_{B_s} - M_B$ is then the value at 
$M_{\eta_s}$ = 0.6858 GeV and $M_{\pi^0}$ = 0.135 GeV, for 
$a=\delta x_l = \delta x_s = 0$. We obtain 0.073(14) GeV.    
This value is to be compared to the experimental 
mass difference between the $B_s$ meson and the average 
of the charged and neutral $B$ mesons which we denote 
$B_l$ (thus averaging the 
$u$ and $d$ quark masses). However it has to be adjusted 
for electromagnetic effects not included in our lattice 
QCD calculation. Following the discussion in section~\ref{subse:bs}
we see that electromagnetic effects in the $B_s$ and 
$B_d$ mesons are very small. For the $B_u$ however, because 
it is charged, the shift is more substantial at 2.2 MeV 
\footnote{The charged and neutral B mesons are experimentally very close 
in mass but this is the result of electromagnetic 
and $u/d$ mass difference effects that largely cancel. Here $u/d$ mass 
differences are not relevant because we calculate the $u/d$ average 
but we still have to take electromagnetic effects into account.}.  
Adding in electromagnetism then shifts our $M_{B_s}-M_{B_l}$ 
splitting down by 1 MeV. The result 72(14) MeV is shown as 
the shaded green band in Figure~\ref{fig:bsbplot}. It 
is in reasonable agreement with the experimental result 
of 87 MeV~\cite{pdg}. 

Our final result for $M_{B_l} = (M_{B^\pm}+M_{B^0})/2$ is 
5.363 - 0.072 = 5.291(11)(14) GeV. The first error comes 
from the mass of the $B_s$ and is discussed in 
subsection~\ref{subse:bs}, the second comes from 
the mass difference between the $B_s$ and $B_l$. We 
do not expect any significant additional systematic 
errors from NRQCD, beyond those that the $B_l$ inherits 
from the $B_s$ in this method, because those errors 
should cancel in $M_{B_s}-M_{B_l}$. The error budget 
for $M_{B_l}$ is then as given for $M_{B_s}$ 
in Table~\ref{tab:errors} 
with the additional 14 MeV given above.  
Our result for $M_{B_l}$ of 5.291(18) GeV can be compared to the 
experimental result of 5.2794(3) GeV~\cite{pdg}.

\subsection{$B^*-B$ splittings}
\label{se:hyp}
By projecting out the vector states at the source and sink we can measure 
the correlator for the $B^*$, $B^*_s$, and $B^*_c$. As they come from exactly
the same configurations and valence HISQ propagators as the corresponding 
pseudoscalar states, they are highly correlated. In this case 
we can do simultaneous fits to both the pseudoscalar and vector meson 
propagators, and extract a value for the $M_{B^*} - M_{B}$ splittings.

This hyperfine splitting is generated 
by the $\sigma\cdot B$ term in 
the NRQCD action, equation~\ref{deltaH}. This is $\cal{O}$$(v_b^4)$ in 
the relativistic power counting for heavyonium and 
$\cal{O}$$(\Lambda/M_b)$ in heavy-light power counting. 
In our action we take the coefficient 
of this term, $c_4$ to be 1, but it will have radiative corrections 
when matched through $\cal{O}$$(\alpha_s)$ with full QCD. 
We are also missing higher dimension operators that correct 
for discretisation errors and add relativistic corrections. 
For $B$ systems, which are relatively large with very slow-moving 
$b$ quarks, we do not expect these latter effects to be 
as important as the issue of the determination of $c_4$ beyond 
tree level. The heavy-light hyperfine splitting generated by 
the $\sigma\cdot B$ term is proportional to $c_4$ 
and so uncertainty in 
$c_4$ leads directly to an 
$\cal{O}$$(\alpha_s)$ i.e 25\%, uncertainty in the splitting which 
decreases only slowly on finer lattices. 
Thus to determine this splitting accurately we need a determination 
of $c_4$. 

Since we use exactly the same NRQCD action for all our calculations, 
however, we can effectively determine $c_4$ by comparing 
one set of heavy-light hyperfine splittings to experiment and then predicting 
the others. Equivalently we can take ratios of hyperfine splittings
in which the normalisation factor, $c_4$, cancels. This is what 
we did in~\cite{Gregory:2009hq}. By using the $B_s^*-B_s$ mass 
difference, which is 46.1(1.5) MeV from experiment~\cite{pdg}, 
we showed that this splitting 
does not depend on the mass of the lighter quark even for as 
heavy a quark as the charm quark, 
and we were able to predict a $B_c^*-B_c$ splitting of 53(7) MeV. 

\begin{figure}
\begin{center}
\includegraphics[width=80mm]{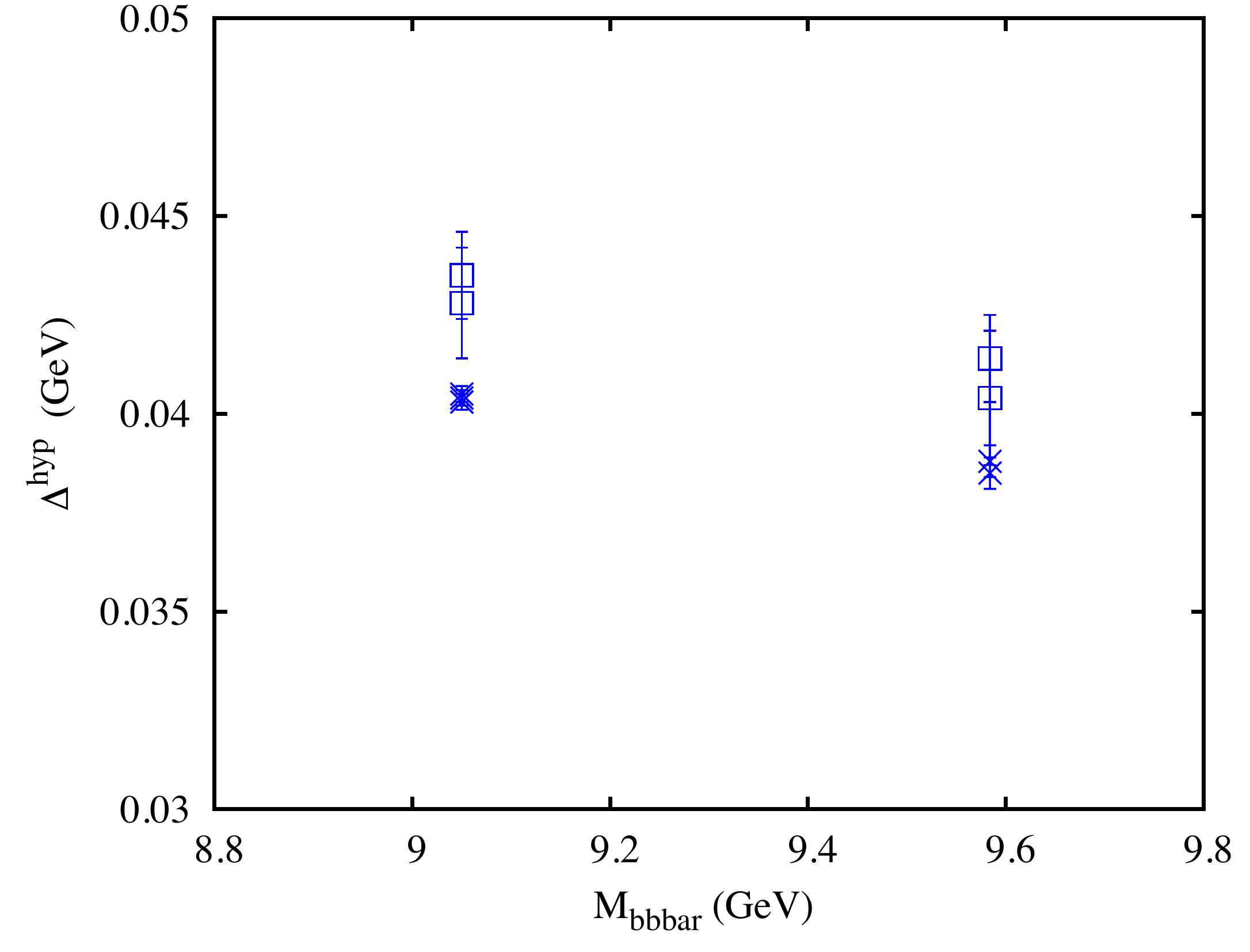}
\end{center}
\caption{Results for the mass difference between 
the $B_s^*$ and $B_s$ mesons (open squares) and 
between $B_c^*$ and $B_c$ mesons (crosses) as a function 
of the spin-average of the $\Upsilon$ and $\eta_b$ 
meson masses used to tune the $b$ quark mass. 
$B_s^*-B_s$ results include two values of the 
$s$ quark mass, and $B_c^* - B_c$ results include 
two different values of the $c$ quark mass as well as
results for $c_1$, $c_5$ and $c_6$ coefficients
differing from 1. 
}
\label{fig:hypvmb}
\end{figure}

We will not discuss that analysis further here, but we give
the table of results of the hyperfine splittings for completeness 
in Tables~\ref{tab:bsresults},~\ref{tab:bcresults},~\ref{tab:bresults}.
They include some additional 
values over those in~\cite{Gregory:2009hq} for the purposes 
of further testing systematic errors. Figure~\ref{fig:hypvmb} shows 
such a test in a plot of the mass difference between 
$B_s^*$ and $B_s$ as a function of $M_{b\overline{b}}$ for 
two different $s$ quark masses (the first 4 rows of entries 
in Table~\ref{tab:bsresults}). Dependence on the $b$ quark 
mass is visible, but dependence on the $s$ quark mass 
is very small. Results are also shown for $B_c^*-B_c$ 
for the same $b$ quark masses and they show a parallel 
slope. In addition we show a result for the the NRQCD 
action with $c_i$ coefficients different from 1 and no 
change is seen. 

\begin{figure}
\begin{center}
\includegraphics[width=80mm]{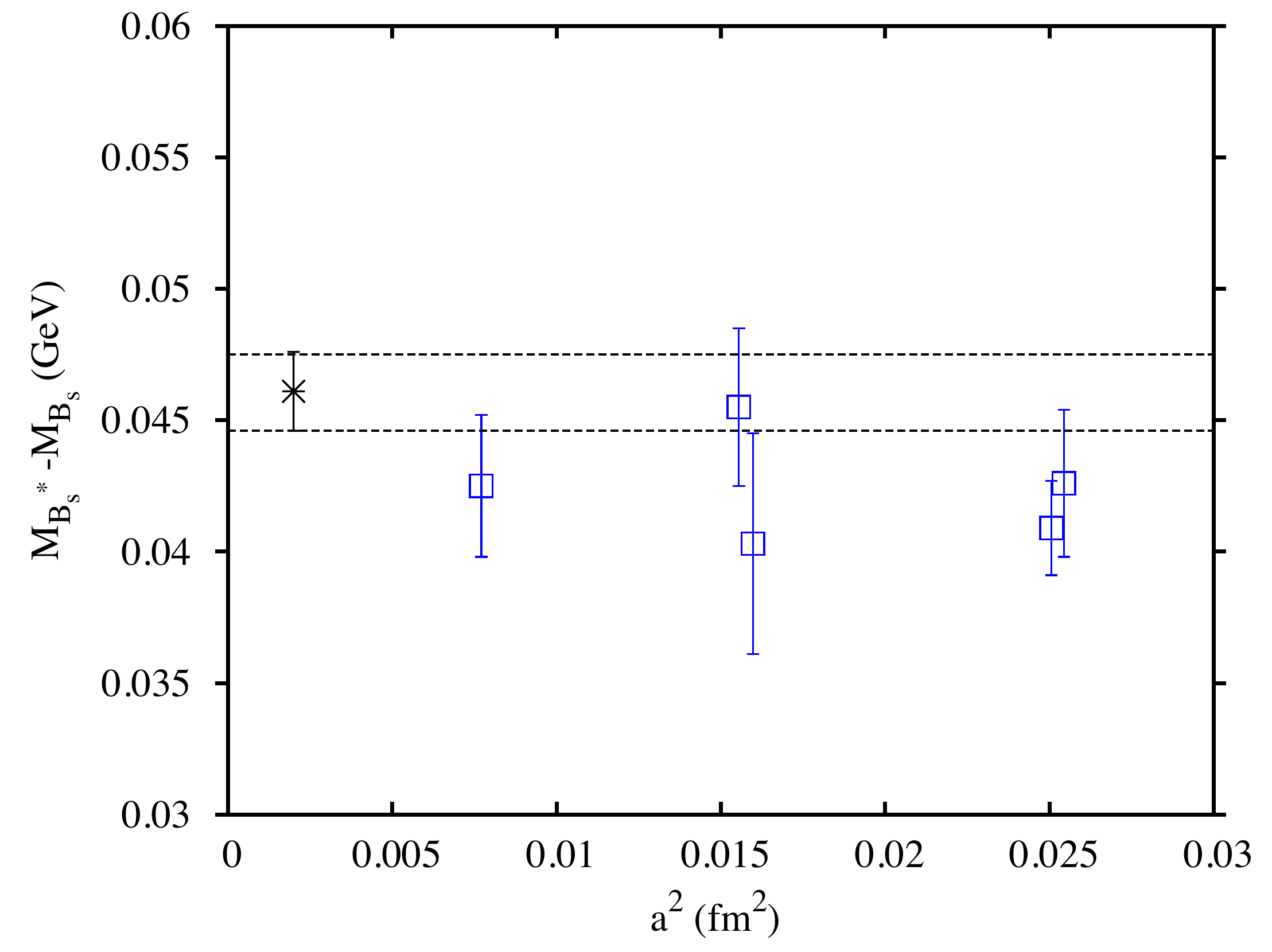}
\end{center}
\caption{Results for the mass difference between 
the $B_s^*$ and $B_s$ mesons as a function 
of lattice spacing.  Results have been corrected 
for mistuning of the $b$ quark mass and the 
errors include statistics, mistuning and lattice 
spacing uncertainties. 
The black star and block dashed lines give the 
experimental result~\cite{pdg}.  
}
\label{fig:hypasq}
\end{figure}

Here we are interested in 
analysing the systematic error in the $B$, $B_s$ and $B_c$ meson masses 
from any uncertainty in $c_4$. We do this by comparing 
our $B_s^*-B_s$ splitting to experiment and interpreting any 
mismatch as a signal for $c_4 \neq 1$. We use the $B_s^*-B_s$ 
because this is the most accurately determined splitting from 
our analysis that is also known experimentally.  
This method can 
be used as a nonperturbative determination of $c_4$, and  
we used this previously to bound the errors on our prediction 
of the hyperfine splitting in bottomonium based on earlier 
$B_s$ and $B$ hyperfine splitting results~\cite{Gray:2005ur}.
Figure~\ref{fig:hypasq} shows our results as a function of 
lattice spacing. We have adjusted them for mistunings of 
the $b$ quark mass according to the results in Figure~\ref{fig:hypvmb} 
but the corresponding shifts are small, and less than the statistical 
errors in all cases. The results show little sign of any lattice 
spacing dependence or sea quark mass dependence 
and we see that a value of $c_4$ of $\approx 1.1$ 
would give agreement with experiment for all the values. 
We therefore estimate that the correct value of $c_4$ for 
this NRQCD is 1.1(1) and that we make a systematic error 
of about 10\% in the heavy-light hyperfine splitting by 
using $c_4 = 1$. This produces a systematic error on 
the $B_c$ and $B_s$ meson masses discussed in 
earlier subsections and included in Table~\ref{tab:errors}. 

Note that the behaviour of the hyperfine splitting in 
bottomonium is quite different from that of 
the $B_s$, being strongly dependent 
on the lattice spacing~\cite{Gray:2005ur}. However, it 
is the same 
operators in the NRQCD action, with the same coefficients, 
that control the fine structure in both systems. 
The matrix elements of the operators can behave quite differently, 
and bottomonium is expected to be a lot more sensitive to 
the lattice spacing than the $B_s$. This means that the $B_s$ 
is a good system from which to determine $c_4$ because it 
is really only sensitive to that coefficient. 

\subsection{Scalar and Axial vector meson masses}
\label{se:scalar}
When generating the NRQCD propagators we choose Dirac structures $\Gamma$ to 
explicitly project out pseudoscalar and vector mesons. 
Parity partners of both of these contribute to their 
correlators, as shown in equation~\ref{eq:fitform}, and 
must be included in the fit. 
The parity 
partner state of the pseudoscalar is a scalar meson and 
the vector meson has as its parity partner a axial-vector state.
So, by carefully fitting the correlators of the $0^-$ and $1^-$ states
we get also the spectra of the $0^+$ and $1^+$ states for free.

In fact our fit results return directly the mass difference between 
the ground state in the oscillating channel and the ground state 
in the non-oscillating channel i.e the $0^+-0^-$ and $1^+-1^-$ mass 
differences. We report these results in Table~\ref{tab:bsresults} for 
the $B_s$ (i.e. for the $B_{s0}^*$ and $B_{s1}$ mesons). 
For $B_l$ and $B_c$ our 
errors are too large on some fits to give a full picture across 
all ensembles. 

\begin{figure}
\begin{center}
\includegraphics[width=80mm]{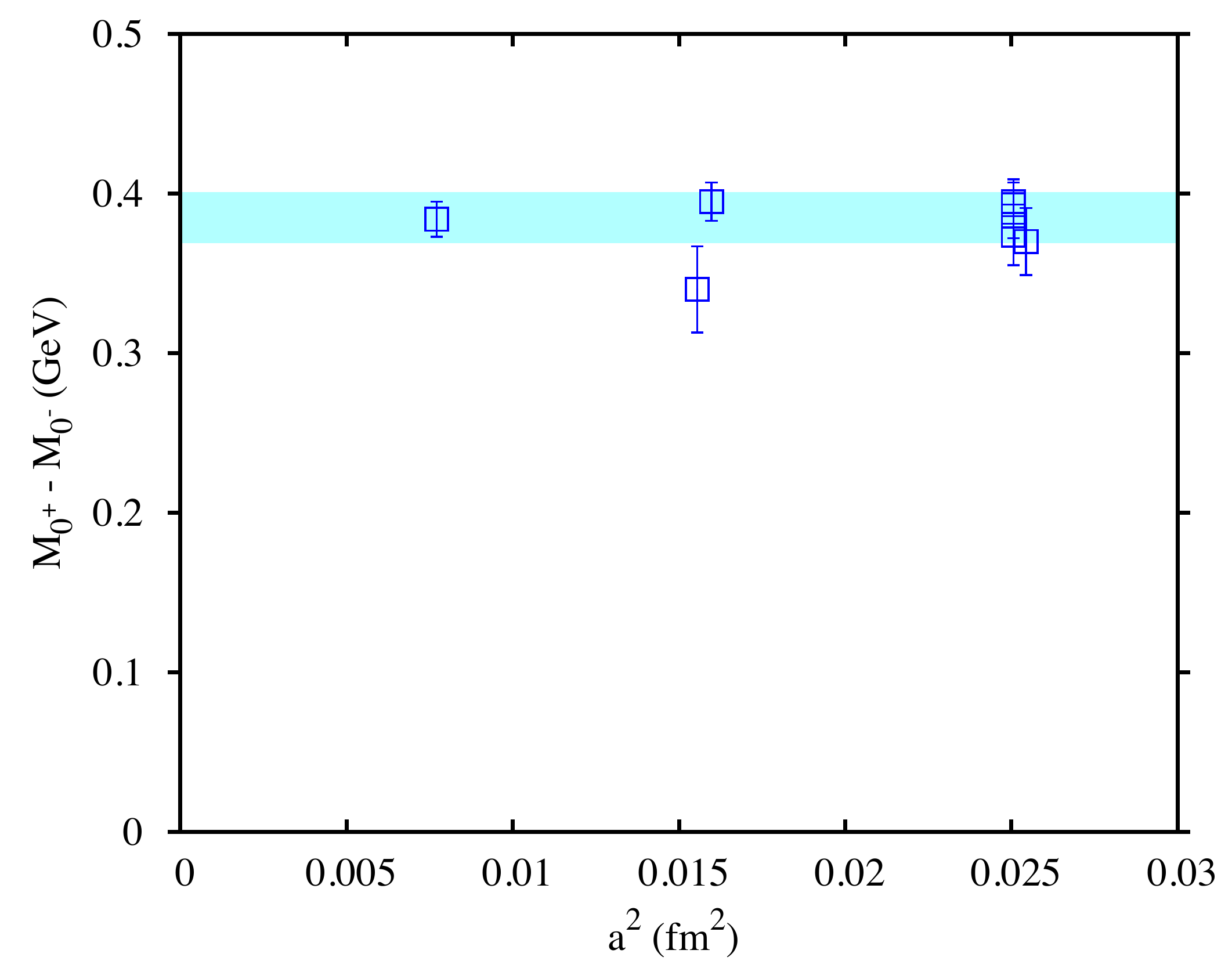}
\end{center}
\caption{Results for the mass difference between 
the scalar $B_{s0}^*$ and $B_s$ mesons as a function 
of lattice spacing. The squares show our results 
(with multiple $b$ and $s$ quark masses on the 
very coarse ensemble, set 1) and the shaded band
the physical result from the fit described in the 
text. This result does not include any adjustment 
or error for the fact that the scalar is not a 
gold-plated meson.  
}
\label{fig:mscalar}
\end{figure}

\begin{figure}
\begin{center}
\includegraphics[width=80mm]{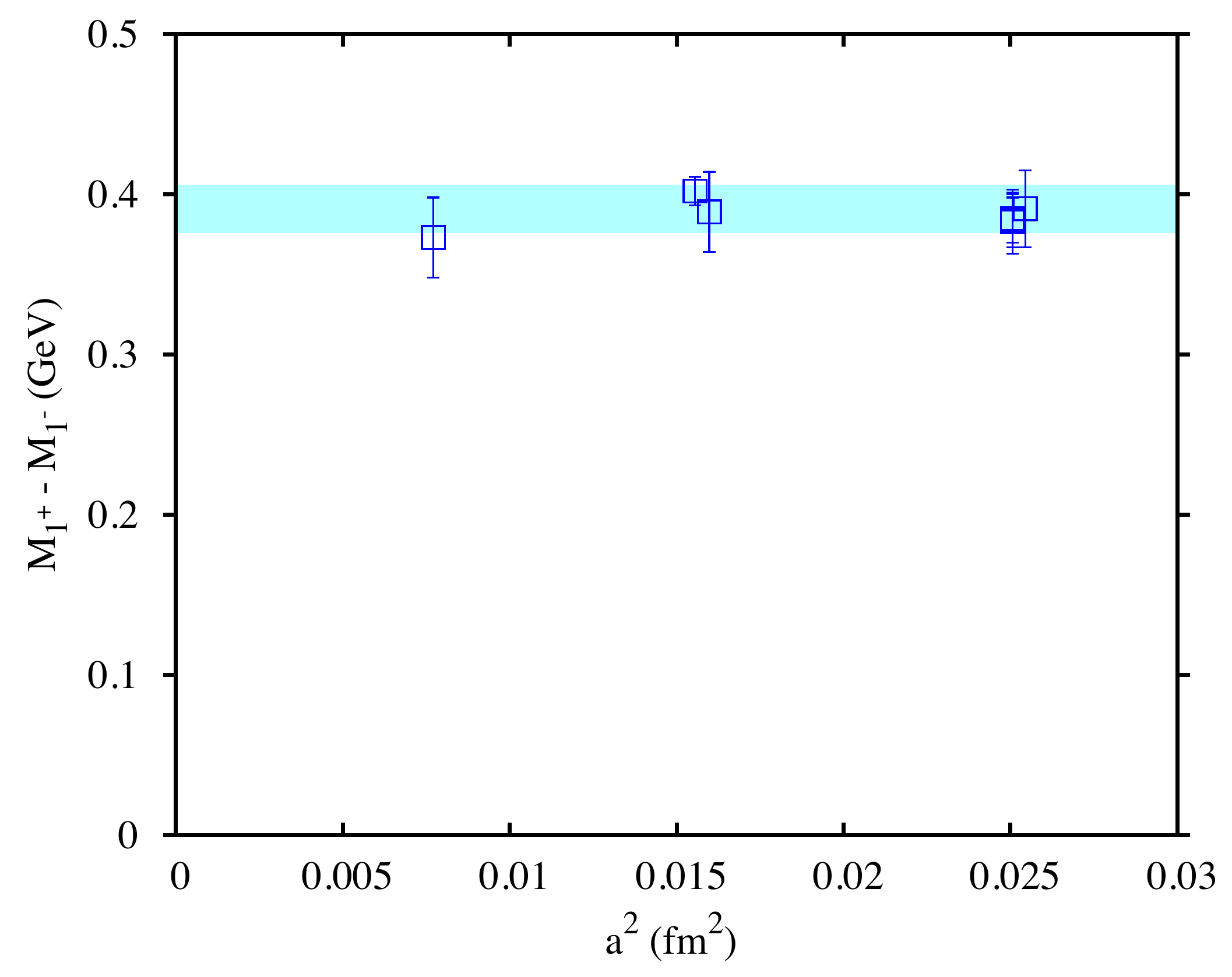}
\end{center}
\caption{Results for the mass difference between 
the axial vector $B_{s1}$ and $B_s^*$ mesons as a function 
of lattice spacing. The squares show our results 
(with multiple $b$ and $s$ quark masses on the 
very coarse ensemble, set 1) and the shaded band
the physical result from the fit described in the 
text. This result does not include any adjustment 
or error for the fact that the axial vector is not a 
gold-plated meson.  
}
\label{fig:maxial}
\end{figure}

The results for the scalar-pseudoscalar mass difference 
are shown in Figure~\ref{fig:mscalar}. There is no signal 
for any systematic dependence on the $b$ or $s$ quark 
mass, or on the lattice spacing. In deriving a physical 
result we allow for both physical and unphysical 
lattice spacing dependence as described for the $B_s$ 
mass in subsection~\ref{subse:bs}, as well as sea
quark mass dependence. We use the same fit form as 
for the $B_s$ mass, given in equation~\ref{eq:fitform}.
The priors are taken to be the same except that we 
take the prior on the physical value of the scalar-pseudoscalar 
mass splitting to be 0.4(2). We also allow for more 
sea quark mass dependence than in that case, because 
the scalar meson is not gold-plated (this will be discussed
further below). We therefore do not take the factor three 
suppression of sea quark mass effects in this case, so 
the prior on the sea quark mass dependent terms is simply 
0.0(2) for the linear terms and 0.00(4) on the quadratic 
terms. The physical result we obtain is 0.385(16) GeV 
and this is plotted as the shaded band on Figure~\ref{fig:mscalar}.  

Exactly the same procedure is followed for the 
axial vector - vector splitting. The results are 
plotted in Figure~\ref{fig:maxial}. From the 
same fit as that described above we obtain 
the physical result for the mass difference 
between the axial vector and vector of 0.391(15) GeV, 
plotted on the Figure 
as a shaded band. 

Since we have calculated the mass differences between 
the scalar and axial vector $B_s$ mesons and the corresponding 
pseudoscalar and vector $B_s$ mesons we expect only 
very small systematic errors coming from NRQCD. Because 
the $b$ quark is very nonrelativistic in these systems, 
as discussed in subsection~\ref{subse:bs} the errors 
from missing higher order relativistic corrections
in NRQCD are very small. They will be reduced further here 
by cancellation in the mass difference. The main 
source of systematic error from NRQCD will come from 
radiative corrections to spin-dependent terms in the 
NRQCD action. In subsection~\ref{se:hyp} we showed 
that these systematic errors are not large, at least 
for the $\sigma\cdot B$ term. There the errors amounted 
to 10\% of the hyperfine splitting, around 5 MeV, split 
between the vector and pseudoscalar states.   
Assuming a similar error for other spin-dependent terms 
which would affect $p$-wave states, we take a systematic 
error of 5 MeV from NRQCD in the mass differences. 

A potentially much larger source of systematic error 
is the fact that the scalar and axial vector mesons 
have strong decay modes, i.e. they are not 
`gold-plated'. This will be discussed further 
in the next section. The strongest decay mode, if 
kinematically allowed, will be to $BK$ (for the $B_{s0}^*$) 
or $B^*K$ (for the $B_{s1}$). If the masses are such 
that the mesons are below threshold for this decay mode, there 
will still in principle be coupling between the meson and 
this virtual decay channel which can shift the meson mass. 
There is in addition a Zweig-suppressed decay mode 
to $B_s\pi$/$B_s^*\pi$ which will be kinematically possible.  

On the lattice the coupling between single and multiparticle 
states is distorted by the fact that $u/d$ quark masses 
are heavier than their physical values and the volume 
of the lattice is relatively small. The fact that $u/d$ masses 
are unrealistic means that decay thresholds are higher than 
in the real world. In principle sensitivity to decay 
thresholds would be seen in the results 
as sea quark mass dependence, but that may not become 
visible until much closer to real world $u/d$ mass values.  
The finite volume of the lattice restricts the decay momenta 
that real or virtual multiparticle states can have. 
A lattice analysis on multiple volumes allows single and 
multiparticle states to be separated. In practice~\cite{Foley:2007ui} 
it seems that bilinear operators of the kind that we have used 
here have very small overlap with multiparticle states.  
So, although in principle there may be a multiparticle 
state (such as $B_s\pi$) at a lower mass value than 
the $B_{s0}$ it is very hard to pick it out of a lattice 
QCD calculation without explicitly using multiparticle 
operators, which we have not done. 
   
A simple model to analyse the effect of multiparticle states is 
one in which point-like meson states are coupled together 
via a perturbation which is a simple 
point-like vertex. We can then calculate the shift on 
the single particle energy from this coupling by 
integrating over the momenta of the decay products in 
the initial particle rest frame. For example, for 
$B_{s0}^*$ coupling to $BK$:
\begin{equation}
\Delta_{E_{B_{s0}^*}} = g^2 \int^{\Lambda} \frac{d^3p}{(2\pi)^3} \frac{1}{M_{B_{s0}^*} - (E_{B} + E_K)}.
\end{equation}
$\Lambda$ represents an ultraviolet cut-off required for this 
model to make sense. $\Lambda \approx$ 500 MeV 
and $g^2 \approx (0.5/\Lambda)$.
If our calculation is correct that the $B_{s0}^*$ is close to, but
below, threshold then we can treat the 
$B$ and $K$ as nonrelativistic and, dropping the $B$ kinetic
term, 
\begin{equation}
\Delta_{E_{B_{s0}^*}} = -\frac{g^2}{2\pi} \int^{\Lambda} \frac{p^2dp}{p^2/(2M_K)+\Delta M}.
\end{equation}
where $\Delta M = M_B+M_K-M_{B_{s0}^*}$ (unperturbed values). 
The $\Delta M$-dependent piece of the mass shift is then 
given by:
\begin{equation}
\Delta_{E_{B_{s0}^*}} = -\frac{g^2 \sqrt{\Delta M 2 M_K} 2 M_K}{4}
\end{equation}
Numerically this gives a shift downwards of a 
few tens of MeV for $\Delta M$ values of 
a few tens of MeV. From this we conclude that a reasonable 
systematic error for the absence of coupling to strong decay channels 
is 25 MeV (which we take to be a symmetric error). 

This then gives the following mass differences: 
\begin{eqnarray}
M_{B_{s0}^*} - M_{B_s} &=&  0.385(16)(5)(25)        \nonumber \\
M_{B_{s1}} - M_{B_s^*} &=&  0.391(15)(5)(25)
\end{eqnarray}
where the first error is statistics/fitting, the second is 
the NRQCD systematic error and the third is the error 
from not including coupling to strong decay channels.  

\section{Discussion}
\label{se:discussion}

Figure~\ref{fig:mbsasq} shows that our result for the mass of 
the $B_s$ meson agrees well with experiment with total errors 
of 11 MeV (0.2\%). The errors are dominated by statistical errors 
and systematic errors from NRQCD, both of which are being improved 
in work underway.  

As discussed earlier, because we fix the $b$ quark and $s$ quark 
masses from other mesons, the $B_s$ mass determination is completely
free from any parameter tuning. An alternative for the $b$ quark 
mass, adopted by some other lattice QCD calculations is to 
fix the $b$ quark mass from the $B_s$ meson mass itself. 
However, it is still possible then to determine 
$\Delta_{B_s} = M_{B_s} - M_{b\overline{b}}/2$, as a test of 
the $b$ quark systematic errors. The only other full lattice 
QCD calculation of this quantity is from the 
Fermilab Lattice/MILC collaboration using the 
Fermilab formalism for the $b$ quark~\cite{fnalheavy09}. They determine in 
fact the quantity 
$\Delta_{\overline{B_s}} = M_{\overline{B_s}} - M_{b\overline{b}}/2$ 
where $\overline{B_s}$ is the spin average of the 
$B_s$ and the $B_s^*$ masses. This quantity has reduced 
systematic errors from the spin-dependent terms  
in the action, in the same way that the use of $M_{b\overline{b}}$ 
reduces these systematic errors for the bottomonium 
system. The Fermilab Lattice/MILC collaboration obtain 
the value $1359 \pm 304 {+31 \atop -0}$ MeV for 
$2\Delta_{\overline{B_s}}$ with a partial error budget~\cite{fnalheavy09}.
We can also determine $\Delta_{\overline{B_s}}$ in exactly 
the same way as we determined $\Delta_{B_s}$. 
We obtain 0.671(7) GeV for the physical result from 
our calculation. This becomes 0.675(11) GeV when 
corrected for electromagnetic, annihilation and 
charm-in-the-sea effects in $b\overline{b}$ and 
with a full error budget (essentially the same 
as in Table~\ref{tab:errors} but with a reduced
error for NRQCD systematics in the $B_s$). 
The experimental result is 0.6817(11) GeV~\cite{pdg}. 

\begin{figure}
\begin{center}
\includegraphics[width=80mm]{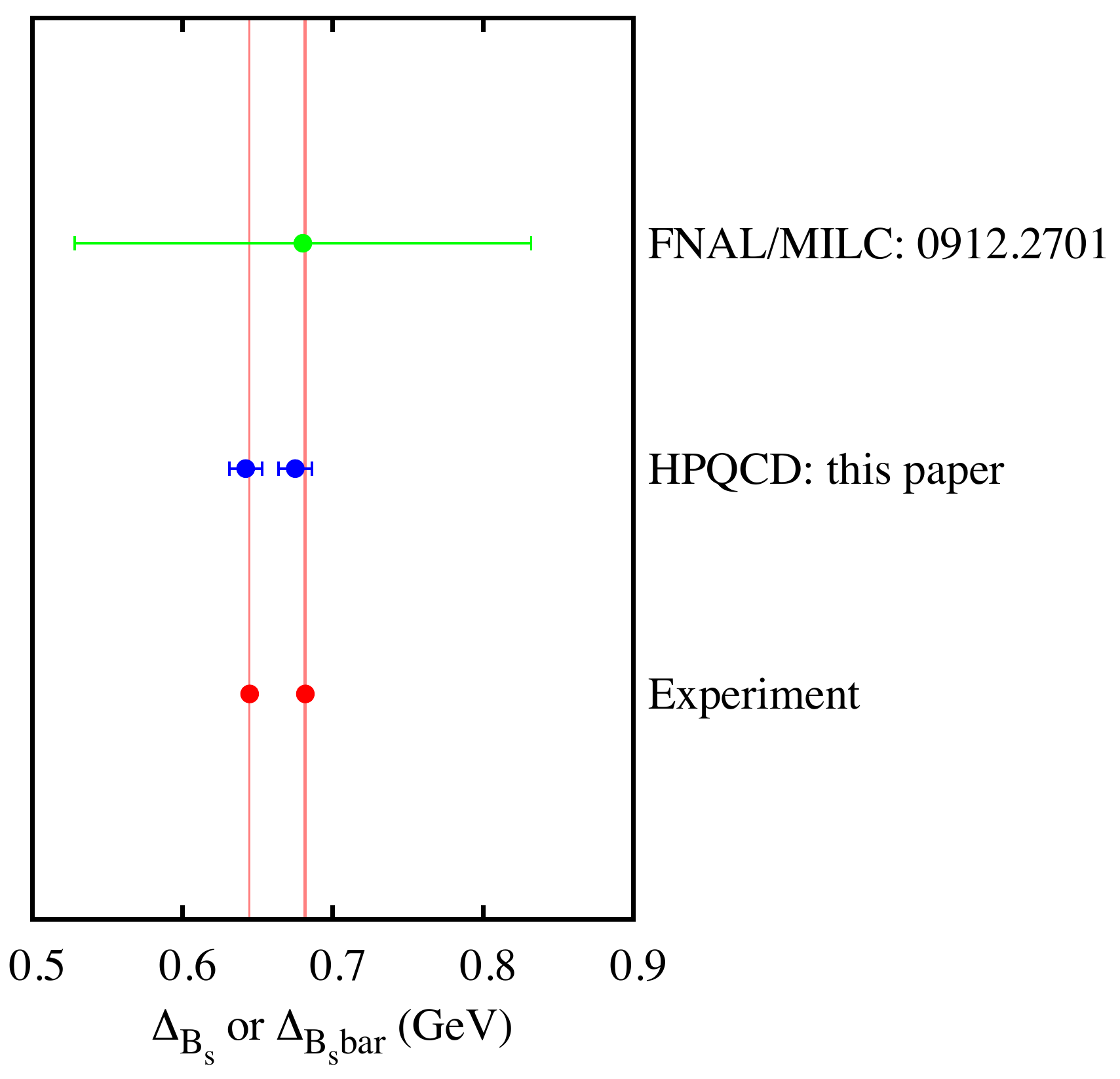}
\end{center}
\caption{Results for the mass differences 
$\Delta_{B_s}$ and $\Delta_{\overline{B_s}}$ 
between the $B_s$ and the spin average of 
$B_s$ and $B_s^*$ respectively and the spin 
average of the $\Upsilon$ and $\eta_b$ (see 
text). The top result is from the Fermilab 
Lattice/MILC collaboration for $\Delta_{\overline{B_s}}$~\cite{fnalheavy09}, 
the middle two results are from this paper, 
and the lower two points, and shaded vertical lines, are from experiment. 
}
\label{fig:summary}
\end{figure}

Figure~\ref{fig:summary} shows the results for 
$\Delta_{B_s}$ and $\Delta_{\overline{B_s}}$ 
from this paper and from the Fermilab Lattice/MILC
collaboration compared to experiment. 
Both results agree with experiment but we are 
able to provide a 2\% test of these mass 
differences, which is a nontrivial test of QCD.  

\begin{figure}
\begin{center}
\includegraphics[width=80mm]{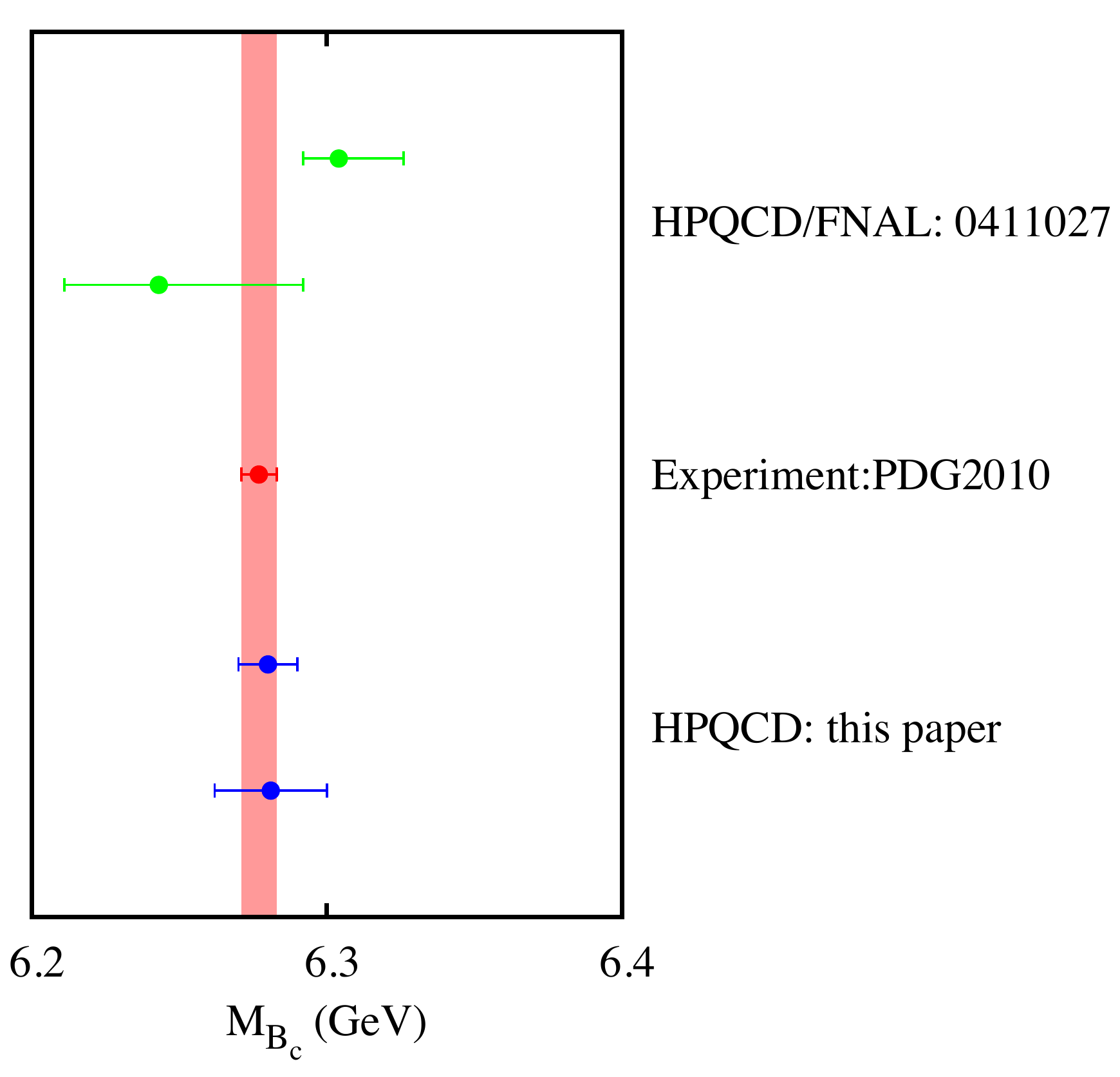}
\end{center}
\caption{Results for the mass of the $B_c$ 
obtained from the hh and hs methods. 
The top two results (hh above hs) are the 2004 HPQCD/Fermilab 
Lattice calculation~\cite{allisonbc} and the bottom two results 
(again hh above hs) are from this paper. 
The middle point, and the shaded vertical line, is the 
current experimental value.   
}
\label{fig:summarybc}
\end{figure}

Back in 2004 we predicted the mass of the $B_c$ ahead 
of the CDF experimental discovery in a collaboration with the Fermilab Lattice 
collaboration~\cite{allisonbc}. We used NRQCD for the $b$ quarks, as here, 
but the Fermilab formalism for the $c$ quarks, and the asqtad 
formalism for the $s$ quarks in the hs method. As a result, we had 
larger statistical and systematic errors than we have here, 
particularly for the hs method. 

Figure~\ref{fig:summarybc} shows the comparison between 
our old results and the new ones given here, as well 
as the current experimental value. The improvements 
in lattice QCD calculations since 2004, including the
development of the HISQ action for $c$ and $s$, give 
us a substantial improvement in errors and consistency 
between the hh and hs methods today.   
 
\begin{figure}
\begin{center}
\includegraphics[width=80mm]{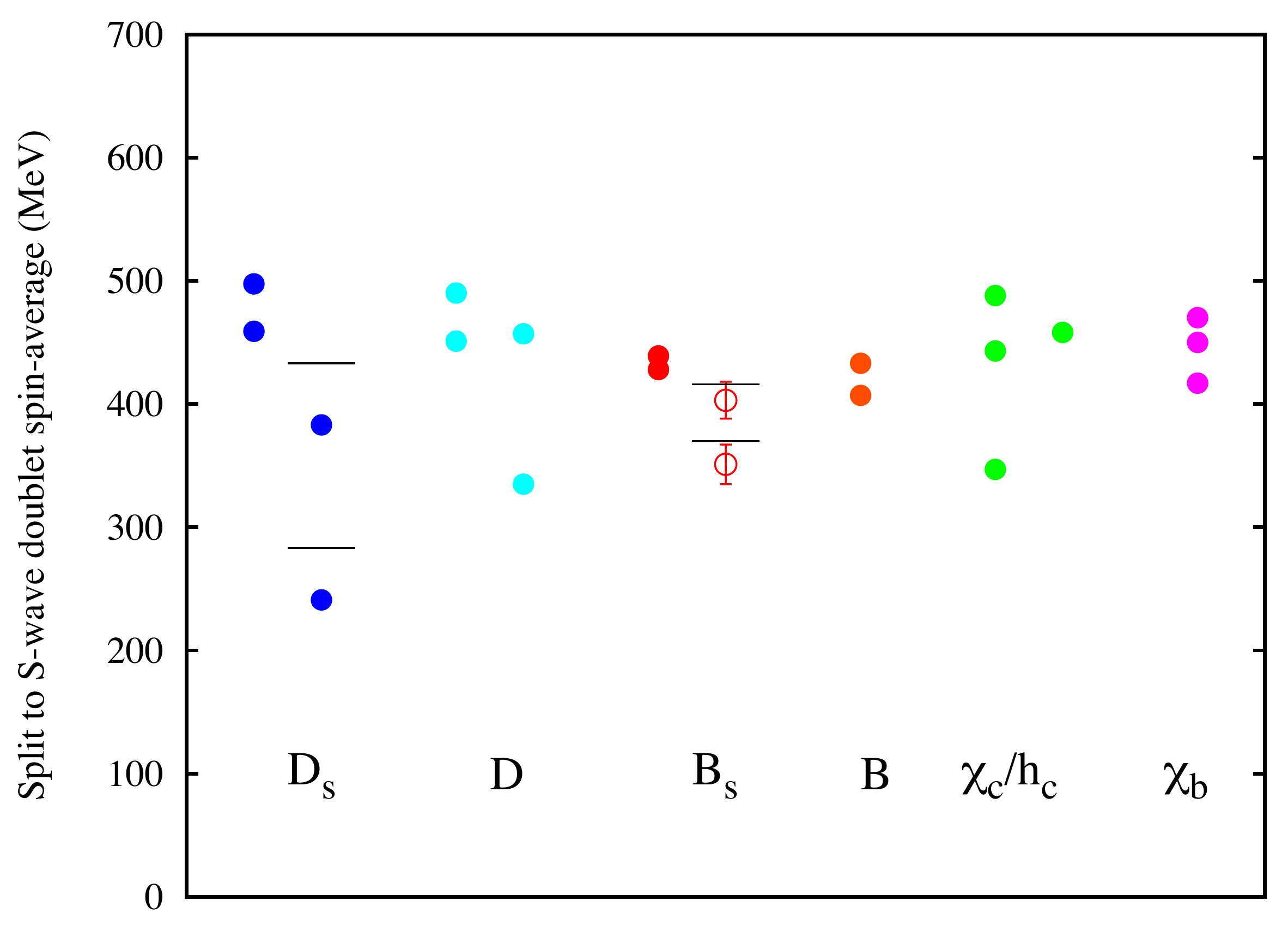}
\end{center}
\caption{Results for the mass difference between 
$0^+$, $1^+$ and $2^+$ $D$, $D_s$, $B$ and $B_s$ 
states and the spin-average of the corresponding 
$0^-$ and $1^-$ states. The solid circles are 
from experiment~\cite{pdg} and are divided, where 
results exist, into the expected $0^+/1^+$ and $1^+/2^+$ 
doublets. For the $B$ and $B_s$ only the 
upper $1^+/2^+$ doublet has been seen. 
The open circles for the $B_s$ are the 
lattice QCD results given here. We have 
not included an error for the coupling 
of these states to 
decay channels. The rightmost two sets 
of points are the $P$-wave charmonium 
and bottomonium states for comparison.   
The black lines show the 
$D^{(*)}K$ and $B^{(*)}K$ thresholds for the Zweig-allowed decay 
of the $0^+$ and $1^+$ $D_s$ and $B_s$ states. 
}
\label{fig:hlpwave}
\end{figure}

Our result for the mass difference between the $B_s$ 
and the $B_l$ meson is the first full lattice QCD 
calculation of this quantity. As discussed in 
subsection~\ref{se:bmass} our result is in agreement 
with experiment, but with substantial statistical 
errors. These will be improved in further work which 
is under way. 

In subsection~\ref{se:scalar} we gave results for 
the masses of the $0^{+}$ and $1^{+}$ $B_s$ mesons. 
The $0^+$ has not been seen experimentally. A $1^+$ state 
has been seen but may not be the one whose mass we 
have calculated.  

Figure~\ref{fig:hlpwave} shows how our results
fit into the current experimental 
picture of `$p$-wave' charm-light and 
bottom-light mesons. Charmonium and bottomonium $p$-wave 
mesons are also shown for comparison~\cite{pdg}. 
For heavy-light mesons the $p$-wave states are expected 
from heavy quark symmetry~\cite{isgurwise} to appear in two doublets, 
classified according 
to the $J$ of the light
quark which can be either 1/2 or 3/2, when $L$=1 is combined 
to $s_l$ = 1/2. 
The $j_l = 1/2$ doublet then separates into 
a $0^+$ and $1^+$ meson at non-infinite heavy quark 
mass, when the heavy-quark spin is coupled in. 
The $j_l = 3/2$ doublet is likewise made up 
of $1^+$ and $2^+$ states. 
This is in contrast 
to the heavyonium case where there is a triplet 
of $0^+$, $1^+$ and $2^+$ states with total quark-antiquark
spin of 1, and a single $1^+$ state with total spin 1.  
The existing experimental results are shown as solid points 
in Figure~\ref{fig:hlpwave} divided appropriately according 
to the picture above. For $D$ and $D_s$ mesons both doublets 
have been seen; for $B$ and $B_s$ mesons only the $j_l = 3/2$ 
doublet has been seen (assuming that the $1^+$ state seen is associated 
with that doublet). For charmonium all 4 states of the $\chi_c$ 
triplet and the $h_c$ are known; for bottomonium the $h_b$ 
has not been seen. The experimental masses are given relative 
to the spin-average of the $s$-wave states, a pseudoscalar and 
a vector in every case. That removes the overall mass scale of each system 
from the plot and shows, as is well-known but still somewhat surprising, 
 that the orbital excitation energies 
of heavy degrees of freedom in heavyonium are very similar
to those of light degrees of freedom in a heavy-light system. 

Since mass splittings between the $S=1$ states in heavyonium 
and between the members of the $j_l=1/2$ or $j_l=3/2$ doublets 
in the heavy-light case 
are caused by heavy quark spin effects proportional to the inverse 
of the heavy quark mass we expect to see larger splittings in 
the $c$ case than in the $b$ case. This is borne out in the 
experimental data for charmonium and bottomonium and in the 
comparison of $D$ and $B$ results for the $j_l$ = 3/2 doublet 
(although the disagreement between $B$ and $B_s$ might 
indicate that the doublet assignment for the $B$ $1^+$ in the 
Figure is wrong).
The splitting between $j_l = 1/2$ and $3/2$ doublets 
is a light quark effect that does not vanish as $m_Q \rightarrow 
\infty$. However the splitting will vary with $m_Q$ slightly 
because of $\Lambda/m_Q$ terms in the effective heavy quark 
action (NRQCD) away from that limit. A variation in the splitting 
of order 100 MeV out 
of 500 MeV is then reasonable between $D$ and $B$.   

Our results are entered on Figure~\ref{fig:hlpwave} as open 
circles in the $B_s$ column. Since we have $0^+$ and $1^+$ 
states we have placed them as the $j_l = 1/2$ doublet. 
However, it should be stressed that we do not know that 
that assignment for the $1^+$ is correct. In any case the $1^+$ states 
from the two doublets can mix and we have not allowed for that. 

Given the discussion above, our results fit fairly naturally 
into the picture described. As a $j_l =1/2$ doublet they sit 
below the known $j_l =3/2$ doublet. They sit closer 
to the $j_l = 3/2$ doublet than for the $D_s$ case, but this can 
be a $\Lambda/m_Q$ effect as discussed above. The splitting between 
the two states in the doublet is about one third that in the 
$D_s$ case, consistent with this splitting being a $1/m_Q$ effect.  

In the $D_s$ case the discovery of the lower $0^+/1^+$ doublet~\cite{babar-ds0, cleo-ds1, belle-ds1} 
caused much surprise because the states were low compared to
model calculations. The states had been expected to be 
above threshold for strong decay to $DK$ and $D^*K$ respectively 
and therefore broad (unlike the upper doublet which has to decay 
in a d-wave). Instead they are below threshold and so decay to 
the Zweig-suppressed $D_s \pi$ and $D_s^* \pi$ channels and are narrow.  
We have marked the $DK$, $D^*K$, $BK$ and $B^*K$ 
thresholds on Figure~\ref{fig:hlpwave}. 
Like the $D^*_{s0}$ our $0^+$ $B_s$ 
state is also below, but very close to, its Zweig-allowed decay threshold. 
A similar situation holds for the $1^+$ state. 
This might indicate that these states would be narrow. However, 
there will also be effects from coupling to the decay 
channel that are not included in our calculation. On our lattices 
the light sea quark masses are heavier than in the real world and 
hence the $K$ mesons containing a valence $s$ quark and a sea light quark
would be too heavy to allow the $B_{s0}^*$ to decay to $BK$. 
$B_s\pi$ decay is allowed but with a very restricted phase space 
compared to the real world. Coupling to these channels would in 
principle show up as sea quark mass dependence but would need a 
bigger range of sea quark masses than we have used.   
In subsection~\ref{se:scalar} we allowed a 25 MeV systematic 
error for these coupled-channel effects, noting that the coupling 
to $BK$ decay will tend to push the mass down. 

Our results are the only ones in full lattice QCD to date and 
with realistic $b$ quarks. 
There have, however, been several 
recent lattice QCD results for the case of $u$ and $d$ sea quarks only 
and taking $b$ quarks in the static 
limit~\cite{Foley:2007ui, Koponen:2007nr, Burch:2008qx, etmcstatic}. 
The most complete is that of the ETMC collaboration~\cite{etmcstatic}. 
They give a mass difference 
between the $B_s$ $j_l=1/2$ doublet and the spin-average of 
$B_s$ and $B_s^*$ of 413(12) MeV, with an estimated additional 
possible systematic error of 20 MeV, including coupling to 
multiparticle states. Using experimental results 
from charmed mesons to estimate $1/m_b$ corrections to the static 
limit, they conclude, as we have done, that the scalar $B_s$ 
state is close to the $BK$ threshold.   

\section{Conclusions}  \label{se:conclusions}
We have given the first accurate result for the $B_s$ meson mass from lattice 
QCD including the effect of $u$, $d$ and $s$ sea quarks, and with a full error 
budget. We have improved significantly on an 
earlier value for the $B_c$ meson mass, achieving smaller errors and better 
consistency between two different methods. 
The determination of both of these masses provides a strong test of 
our lattice QCD approach to $b$ physics, because they test the consistency 
of heavyonium and heavy-strange or heavy-charm physics from the 
same heavy quark action. 
All of the QCD parameters used here are 
tuned from other calculations so our results are parameter free tests of 
QCD against experiment.  

The mass of the $B$ meson, specifically 
the difference between $B_s$ and $B$ meson masses, depends 
on light quark physics since heavy quark effects 
cancel.  Our result agrees with experiment but needs higher 
statistical precision for a good test. 

We also discuss scalar and axial vector meson masses for 
the $B_s$.  Our results indicate masses below, but 
close to, threshold for Zweig-allowed decay modes.  
From our current calculation, however, it is not possible to 
include effects of coupling to either allowed or suppressed
decay channels, so significant shifts to our results from these effects 
are possible. 

Further improvement to these results will come with improved 
statistical accuracy in calculations now underway. This 
will lead also to improved determination of decay constants and 
other $B$ meson matrix elements. Confidence in those calculations 
and the error analysis associated with them is strongly bolstered
by this analysis of the associated meson masses.

\section*{ACKNOWLEDGEMENTS}

We are grateful to Craig McNeile and Rachel Dowdall for useful discussions. 
This work was funded by STFC, the Scottish Universities Physics 
Alliance, NSF, DoE, MICINN, the Cyprus Research Promotion Foundation 
and the EU as part of STRONGNET. 
We thank the MILC collaboration 
for the use of their gluon configurations. 
The computing was done on Scotgrid, the QCDOCX cluster of the 
UKQCD collaboration, NERSC,
 the Ohio Supercomputer Centre and at Fermilab on facilities 
of the USQCD collaboration. 

\bibliographystyle{h-physrev2}
\bibliography{bmes}

\begin{thebibliography}{10}

\bibitem{elvirabb}
HPQCD, E.~Gamiz, C.~T.~H. Davies, G.~P. Lepage, J.~Shigemitsu, and M.~Wingate,
\newblock Phys. Rev. {\bf D80}, 014503 (2009), 0902.1815,
\newblock 

\bibitem{lepage:1992tx}
G.~P. Lepage, L.~Magnea, C.~Nakhleh, U.~Magnea, and K.~Hornbostel,
\newblock Phys. Rev. {\bf D46}, 4052 (1992), hep-lat/9205007,
\newblock 

\bibitem{Follana:2006rc}
HPQCD, E.~Follana {\em et~al.},
\newblock Phys. Rev. {\bf D75}, 054502 (2007), hep-lat/0610092,
\newblock 

\bibitem{Bazavov:2009bb}
A.~Bazavov {\em et~al.},
\newblock Rev. Mod. Phys. {\bf 82}, 1349 (2010), 0903.3598,
\newblock 

\bibitem{Orginos:1998ue}
MILC, K.~Orginos and D.~Toussaint,
\newblock Phys. Rev. {\bf D59}, 014501 (1999), hep-lat/9805009,
\newblock 

\bibitem{Orginos:1999cr}
MILC, K.~Orginos, D.~Toussaint, and R.~L. Sugar,
\newblock Phys. Rev. {\bf D60}, 054503 (1999), hep-lat/9903032,
\newblock 

\bibitem{Lepage:1998vj}
G.~P. Lepage,
\newblock Phys. Rev. {\bf D59}, 074502 (1999), hep-lat/9809157,
\newblock 

\bibitem{horgan}
Z.~Hao, G.~M. von Hippel, R.~R. Horgan, Q.~J. Mason, and H.~D. Trottier,
\newblock Phys. Rev. {\bf D76}, 034507 (2007), 0705.4660,
\newblock 

\bibitem{kendall}
HPQCD, C.~T.~H. Davies, E.~Follana, I.~D. Kendall, G.~P. Lepage, and
  C.~McNeile,
\newblock Phys. Rev. {\bf D81}, 034506 (2010), 0910.1229,
\newblock 

\bibitem{Follana:2007uv}
HPQCD, E.~Follana, C.~T.~H. Davies, G.~P. Lepage, and J.~Shigemitsu,
\newblock Phys. Rev. Lett. {\bf 100}, 062002 (2008), 0706.1726,
\newblock 

\bibitem{Gray:2005ur}
HPQCD, A.~Gray {\em et~al.},
\newblock Phys. Rev. {\bf D72}, 094507 (2005), hep-lat/0507013,
\newblock 

\bibitem{newfds}
HPQCD, C.~T.~H. Davies {\em et~al.},
\newblock (2010), 1008.4018,
\newblock 

\bibitem{schladming}
C.~Davies,
\newblock (1997), hep-ph/9710394,
\newblock 

\bibitem{wingate}
HPQCD, M.~Wingate, J.~Shigemitsu, C.~T.~H. Davies, G.~P. Lepage, and H.~D.
  Trottier,
\newblock Phys. Rev. {\bf D67}, 054505 (2003), hep-lat/0211014,
\newblock 

\bibitem{oldups}
C.~T.~H. Davies {\em et~al.},
\newblock Phys. Rev. {\bf D50}, 6963 (1994), hep-lat/9406017,
\newblock 

\bibitem{eike}
E.~Muller,
\newblock Phd thesis, University of Edinburgh  (2009).

\bibitem{hpqcdinprep}
HPQCD,
\newblock in preparation .

\bibitem{exptetab}
BABAR, B.~Aubert {\em et~al.},
\newblock Phys. Rev. Lett. {\bf 101}, 071801 (2008), 0807.1086,
\newblock 

\bibitem{mbmcpaper}
HPQCD, C.~McNeile, C.~T.~H. Davies, E.~Follana, K.~Hornbostel, and G.~P.
  Lepage,
\newblock Phys. Rev. {\bf D82}, 034512 (2010), 1004.4285,
\newblock 

\bibitem{pdg}
Particle Data Group, K.~Nakamura,
\newblock J. Phys. {\bf G37}, 075021 (2010),
\newblock 

\bibitem{gplbayes}
G.~P. Lepage {\em et~al.},
\newblock Nucl. Phys. Proc. Suppl. {\bf 106}, 12 (2002), hep-lat/0110175,
\newblock 

\bibitem{Gregory:2009hq}
HPQCD, E.~B. Gregory {\em et~al.},
\newblock Phys. Rev. Lett. {\bf 104}, 022001 (2010), 0909.4462,
\newblock 

\bibitem{ericlat09}
E.~B. Gregory {\em et~al.},
\newblock PoS(LAT2009)092  (2009), 0911.2133,
\newblock 

\bibitem{morning}
C.~J. Morningstar,
\newblock Phys. Rev. {\bf D50}, 5902 (1994), hep-lat/9406002,
\newblock 

\bibitem{fnalheavy09}
T.~Burch {\em et~al.},
\newblock Phys. Rev. {\bf D81}, 034508 (2010), 0912.2701,
\newblock 

\bibitem{allisonbc}
HPQCD/Fermilab Lattice, I.~F. Allison {\em et~al.},
\newblock Phys. Rev. Lett. {\bf 94}, 172001 (2005), hep-lat/0411027,
\newblock 

\bibitem{isgurwise}
N.~Isgur and M.~B. Wise,
\newblock Phys. Rev. Lett. {\bf 66}, 1130 (1991),
\newblock 

\bibitem{babar-ds0}
BABAR, B.~Aubert {\em et~al.},
\newblock Phys. Rev. Lett. {\bf 90}, 242001 (2003), hep-ex/0304021,
\newblock 

\bibitem{cleo-ds1}
CLEO, D.~Besson {\em et~al.},
\newblock Phys. Rev. {\bf D68}, 032002 (2003), hep-ex/0305100,
\newblock 

\bibitem{belle-ds1}
Belle, P.~Krokovny {\em et~al.},
\newblock Phys. Rev. Lett. {\bf 91}, 262002 (2003), hep-ex/0308019,
\newblock 

\bibitem{Foley:2007ui}
J.~Foley, A.~O'Cais, M.~Peardon, and S.~M. Ryan,
\newblock Phys. Rev. {\bf D75}, 094503 (2007), hep-lat/0702010,
\newblock 

\bibitem{Koponen:2007nr}
UKQCD, J.~Koponen,
\newblock Phys. Rev. {\bf D78}, 074509 (2008), 0708.2807,
\newblock 

\bibitem{Burch:2008qx}
T.~Burch, C.~Hagen, C.~B. Lang, M.~Limmer, and A.~Schafer,
\newblock Phys. Rev. {\bf D79}, 014504 (2009), 0809.1103,
\newblock 

\bibitem{etmcstatic}
ETM, C.~Michael, A.~Shindler, and M.~Wagner,
\newblock JHEP {\bf 08}, 009 (2010), 1004.4235,
\newblock 

\bibitem{gpllat91}
G.~P. Lepage,
\newblock Nucl. Phys. Proc. Suppl. {\bf 26}, 45 (1992),
\newblock 

\bibitem{sommer}
ALPHA, M.~Della~Morte {\em et~al.},
\newblock Phys. Lett. {\bf B581}, 93 (2004), hep-lat/0307021,
\newblock 

\end{thebibliography}

\end{document}